\documentclass[aps,pre,floatfix,twocolumn,groupedaddress]{revtex4-1}
\usepackage[dvips]{graphicx}
\usepackage{graphicx}
\usepackage{amsmath}
\usepackage{balance}
\usepackage{verbatim}
\usepackage[section]{placeins}
\usepackage{float}

\usepackage[T1]{fontenc}
\usepackage{tikz-cd}
\usetikzlibrary{matrix}
\usetikzlibrary{patterns}
\usetikzlibrary{decorations.pathreplacing}
\usetikzlibrary{shadings}
\usepackage{pgfplots}
\pgfplotsset{compat=newest}
\usepackage{pstricks}    
\usepackage{pgf}
\usepackage{tikz}
\usetikzlibrary{arrows, automata}
\usetikzlibrary{shapes.misc}
\usepackage[inline]{enumitem}

\newcommand{\la}{\left\langle}
\newcommand{\ra}{\right\rangle}

\begin{document}

\title{Folding of Free Polymers Under Confinement by Responsive Hydrogels}

\author{Mahesh Aryal and Alan R.~Denton}
\email[]{alan.denton@ndsu.edu}
\affiliation{Department of Physics, North Dakota State University, Fargo, ND 58108-6050, USA}

\begin{abstract}
Controlling the folding of biopolymers in crowded and confined environments, such as the interior of biological cells, remains a significant challenge. Macromolecular crowding and spatial confinement can strongly influence transition pathways between folded and unfolded states of proteins. Such environments can be mimicked in experiments by confining polymers within soft, tunable media, such as hydrogels – cross-linked polymer networks that swell by absorbing solvent and exhibit stimulus-dependent responsiveness. Here, we develop a self-consistent mean-field theory for the equilibrium swelling and internal composition of a hydrogel immersed in a solution of free polymer coils. The theory extends the Flory-Rehner framework by incorporating free polymer coils, modeled as fluctuating ellipsoids, whose sizes and shapes are characterized by the eigenvalues of the gyration tensor. Confinement of the coils within gel pores is characterized by an effective free-volume fraction, parametrized from numerical calculations for self-avoiding-walk chains between hard walls. Coupled chemical potential equilibrium conditions are solved numerically over ranges of coil molecular weight and free polymer concentration. We find that free polymers drive a non-monotonic swelling response of hydrogels, with osmotic swelling at low concentrations followed by deswelling at higher concentrations, in qualitative agreement with experiment. We further demonstrate that confinement within gel pores reduces the radius of gyration and asphericity of embedded polymer coils, driving them toward compact, globular conformations analogous to folded protein states. These results establish a theoretical foundation for using responsive hydrogels as tunable, artificial chaperone-like systems capable of inducing and controlling biopolymer folding.
\end{abstract}

\maketitle

%%%%%%%%%%%%%%%%%%%%%%%%%%%%%%%%%%%%%%%%%%%%%%%%%%%%%%%%%%%%%%%%

\section{Introduction}\label{intro}

Understanding how biopolymers, such as proteins, RNA, and DNA, respond to confinement is fundamental to deciphering their structure and function in both biological and synthetic systems. Inside biological cells, macromolecules inhabit a very crowded environment. Up to 40\% of the cellular volume is occupied by proteins, nucleic acids, ribosomes, and other macromolecular species~\cite{Ellis2001,Zimmerman1991,hancock2012,denton-cmb2013}. Crowding profoundly influences the thermodynamic stability, folding pathways, and association equilibria of proteins~\cite{Minton2008,Zhou2008review,Cheung2005}. Molecular chaperones, such as the GroEL/ES chaperonin complex, act as nanoscale confinement chambers that encapsulate unfolded proteins within cavities of roughly 4-8~nm diameter, restricting their conformational freedom and promoting productive folding toward the native state~\cite{Koldewey2017,Hartl2002}. The ribosomal exit tunnel similarly confines nascent polypeptide chains during co-translational folding~\cite{Ziv2005}. 

Theoretical analyses by Zhou and Dill~\cite{ZhouDill2001} predicted that confinement within a spherical cage can stabilize the folded state of a protein by more than $10\,k_BT$, an effect arising purely from excluded-volume interactions between the unfolded chain and the confining walls. Subsequent simulation studies by Ping~\textit{et al.}~\cite{Ping2004} and others~\cite{Klimov2002,Zhou2008} have confirmed that both folding and unfolding times increase with confinement, while equilibrium is shifted toward the folded state. These findings underscore the profound role of spatial restriction in regulating protein conformations \textit{in vivo}.

It is important to distinguish between two related but physically distinct phenomena: macromolecular crowding and spatial confinement. Crowding refers to volume exclusion by soluble macromolecular cosolutes that are free to rearrange, whereas confinement refers to restriction of a macromolecule within the fixed boundaries of a cavity or pore~\cite{Eggers2001}. While both effects limit the accessible conformations of a polymer, their thermodynamic consequences differ: confinement can produce stabilization effects that are an order of magnitude larger than crowding at comparable volume fractions~\cite{ZhouDill2001,Zhou2004}. 

Minton~\cite{Minton2008,Minton2005} has reviewed how excluded-volume effects from both crowding and confinement influence macromolecular reaction rates and equilibrium, emphasizing that the two mechanisms are thermodynamically distinct. Experimental studies combining confinement and crowding, e.g., involving proteins encapsulated within hydrogel pores in the presence of polymeric crowders, have revealed synergistic, non-additive stabilization effects, indicating that confinement and crowding act in a coupled, non-separable manner~\cite{ross2020combined}.
 
Synthetic hydrogels~\cite{HydrogelBook2012}, cross-linked polymer networks capable of absorbing large volumes of solvent, offer a tunable experimental platform for mimicking the confinement conditions found inside biological cells~\cite{Peppas2006}. Cryo-SEM imaging of gel matrices reveals a porous internal architecture with pore dimensions in the nanometer-to-micrometer range, within which polymer coils or proteins may be embedded~\cite{kaberova2020microscopic}. 

Microgels are microscopic hydrogels that swell and deswell in response to changes in temperature, pH, concentration, and solvent quality~\cite{Pelton2000,MicrogelBook2011,schurtenberger-ZPC2012,Colla2015,plamper2017functional,Karg2019,scheffold2020pathways,Scotti2022review,zaccarelli2024,moncho-jorda2025}. The responsive swelling of microgels is well described by Flory-Rehner theory, which balances mixing entropy of polymer and solvent molecules, enthalpic polymer-solvent interactions, and the elastic free energy of the deformed network~\cite{FloryRehner1943a,FloryRehner1943b,Flory,Erman1986}. Recent critical assessments have confirmed that Flory-Rehner theory provides a quantitatively accurate description of microgel swelling, particularly at high cross-linking densities~\cite{Lopez2017,Friesen2022}.
 
The ability of hydrogels and microgels to undergo reversible, stimulus-controlled swelling and deswelling suggests an intriguing possibility: using responsive gels as active platforms to induce and control the folding and unfolding of embedded biopolymers. As a gel swells by absorbing solvent, its internal pore size increases, relaxing the confinement imposed on embedded polymer chains and favoring extended, unfolded conformations. Conversely, when a gel deswells, as occurs, for example, when the temperature exceeds the volume phase transition temperature or when the osmotic pressure of the external solution is increased, the pore size decreases, tightening confinement and driving embedded polymers toward compact, folded conformations. This mechanism is directly analogous to the biological function of molecular chaperones, which confine unfolded proteins within a restricted cavity to promote folding~\cite{Koldewey2017,Hartl2002}.

A key advantage of hydrogel-based confinement is its external tunability. Adjusting thermodynamic conditions, such as osmotic pressure or solution composition, allows continuously varying the degree of confinement and thereby sweeping through the folding-unfolding transition of the embedded polymer, without relying solely on temperature or pH. This concept motivates the theoretical framework developed in the present work.
 
Considerable theoretical and computational effort has been devoted to understanding the swelling behavior of microgels and their interactions with solvents, nanoparticles, and other macromolecules. Urich and Denton~\cite{Urich2016} developed a coarse-grained model and Monte Carlo simulation framework for bulk suspensions of compressible microgels governed by Flory-Rehner single-particle free energies and Hertzian elastic pair potentials, establishing the methodology that underlies the present study. Subsequent extensions incorporated electrostatic interactions for ionic microgels~\cite{Weyer2018,Alziyadi2023}, concentration-dependent deswelling~\cite{Brito2019}, and the influence of cross-link density and architecture on microgel softness, as revealed by combined small-angle X-ray and neutron scattering experiments~\cite{Scotti2019}. 

In parallel, theoretical modeling of colloid-polymer mixtures, originating with the Asakura-Oosawa depletion model~\cite{Asakura1954,Asakura1958}, has been generalized to account for the shape and conformational fluctuations of polymer depletants. Lim and Denton~\cite{lim-denton-JCP2016} went beyond the standard Asakura-Oosawa model by representing polymers as soft, fluctuating ellipsoids with shapes characterized by the eigenvalue distributions of the gyration tensor, enabling a more realistic description of excluded-volume interactions in colloid-polymer mixtures~\cite{schaefer1999}.
 
Extending from binary to ternary systems, in which a cross-linked gel network coexists with solvent and free polymer chains, introduces qualitatively new physics. Experimentally, Saunders and Vincent~\cite{Saunders1996,Saunders1997a} demonstrated osmotic deswelling of poly(N-isopropylacrylamide) (PNIPAM) and polystyrene (PS) microgels in the presence of free polymer -- PS or poly(ethyleneglycol) (PEG) -- showing that deswelling increases with both molecular weight and concentration of the free polymer when the latter is excluded from the gel interior. Strikingly, Vincent and coworkers~\cite{bradley-vincent2005,routh2006effect} observed a non-monotonic swelling response: at low concentrations, free poly(ethyleneoxide) (PEO) chains permeate the microgel interiors and cause swelling (via increased internal osmotic pressure), whereas at higher concentrations the external osmotic pressure dominates and the microgels deswell. Fern\'{a}ndez-Nieves~\textit{et al.}~\cite{FernandezNieves2003} showed that ionic microgels resist osmotic deswelling below a threshold stress set by the gel charge density. 

On the theoretical side, Bastide \textit{et al.}~\cite{Bastide1981} provided an early treatment of gel deswelling by polymer solutions, and Benhamou~\textit{et al.}~\cite{Benhamou2016} used the random phase approximation to analyze structural properties and phase behavior of cross-linked networks embedded in polymer solutions. Molina~\textit{et al.}~\cite{Molina2005} extended Flory-Rehner theory to a ternary system of partially cross-linked polymers, solvent, and sol macromolecules (free polymers not bound to the network), demonstrating that free chains inside the gel enhance swelling while those in the external bath suppress it. Recently, Takarai~\textit{et al.}~\cite{Takarai2026} established a universal partitioning law for polymer chains entering flexible networks, demonstrating that the classical Ogston model is inadequate for describing partitioning into flexible gels, a finding directly relevant to the free-volume treatment developed in the present work. 

In previous work~\cite{aryal2026response}, we extended Flory-Rehner theory to ternary mixtures of microgels, solvent, and hard-sphere nanoparticle crowders, solving self-consistent equilibrium conditions for solvent and crowder chemical potentials inside and outside the gel, and validating the theory against Monte Carlo simulations. Our results show that nanoparticle crowding promotes deswelling, thereby weakening interparticle correlations, and that, with increasing concentration of crowders, polymeric gels reach a maximum loading capacity dependent on crowder size and polymer cross-link fraction.

\begin{figure}[h]
\begin{center}
\includegraphics[width=0.9\columnwidth,angle=0]{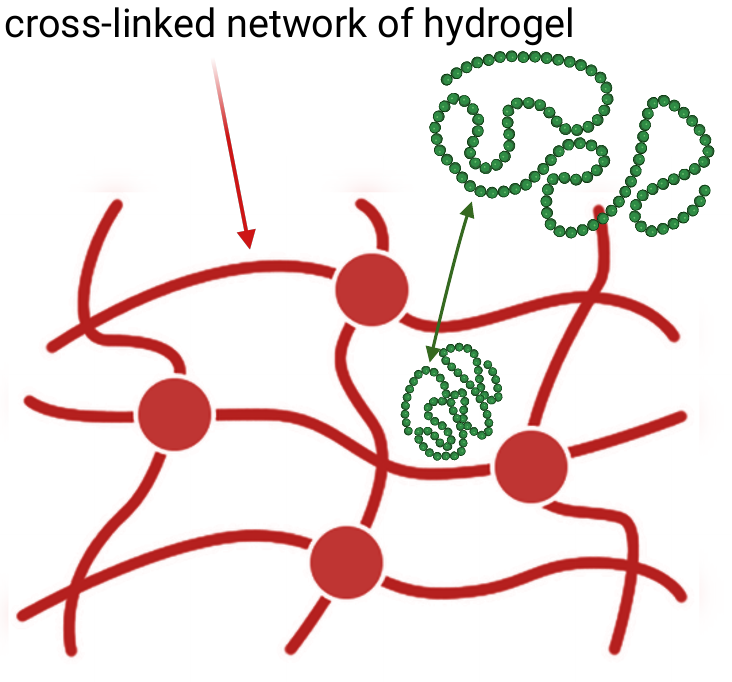}
\end{center}
\caption{Schematic depiction of a hydrogel (red cross-linked network) confining a free polymer coil (green).
Created with BioRender.com/1uo2dyt.}
\label{hydrogelIntake}
\end{figure}

In the present work, we extend this ternary mixture model by replacing rigid, spherical nanoparticle crowders with flexible, self-avoiding random walk (SAW) polymer coils~\cite{deGennes1979,rubinstein2003polymer,Doi2013,Dhar1978}. Each free polymer coil is modeled as a fluctuating ellipsoid~\cite{kuhn1934shape}, whose instantaneous size and shape are characterized by the eigenvalues of the gyration tensor~\cite{Theodorou1985,rudnick1987shapes,Arkin2013,lim-denton-JCP2016}, capturing the full spectrum of conformational fluctuations. The confining geometry of the hydrogel mesh is approximated by parallel hard walls whose separation scales with the swelling ratio. 

The confinement-modified effective free-volume fraction is parametrized from numerical free-volume theory modeling of SAW chains between walls. This approach allows us to self-consistently determine how (i) free polymer coils influence the swelling of the gel through osmotic effects, (ii) the gel selectively absorbs or excludes coils depending on their molecular weight, and (iii) confinement within the gel pores alters the size (radius of gyration) and shape (asphericity) of the embedded coils, driving them toward compact, isotropic conformations reminiscent of folded protein states. 
 
The remainder of the paper is organized as follows. In Sec.~\ref{sec:models}, we outline our coarse-grained
models of swollen hydrogels and free polymer coils and review the Flory-Rehner and free-volume theories. 
In Secs.~\ref{sec:free-energy}, \ref{sec:chempot}, and \ref{sec:computation}, respectively, we identify the free energy 
of the hydrogel-solvent-free polymer mixture, derive the chemical potentials of the free polymer and solvent species,
and outline our computational methods. In Sec.~\ref{sec:results}, we present our numerical results for 
the effective free-volume fraction, radius of gyration, and asphericity of free polymer coils, and for the 
swelling ratio and interior composition of the gel, all as functions of gel pore size and free polymer reservoir 
concentration and molecular weight. Finally, in Sec.~\ref{sec:conclusions}, we summarize and conclude.

\section{Models and Theory}
\label{sec:models}

\subsection{Hydrogel-Solvent-Free Polymer Mixture}
\label{sec:system}

\begin{figure}[h]
\begin{center}
\includegraphics[width=0.9\columnwidth]{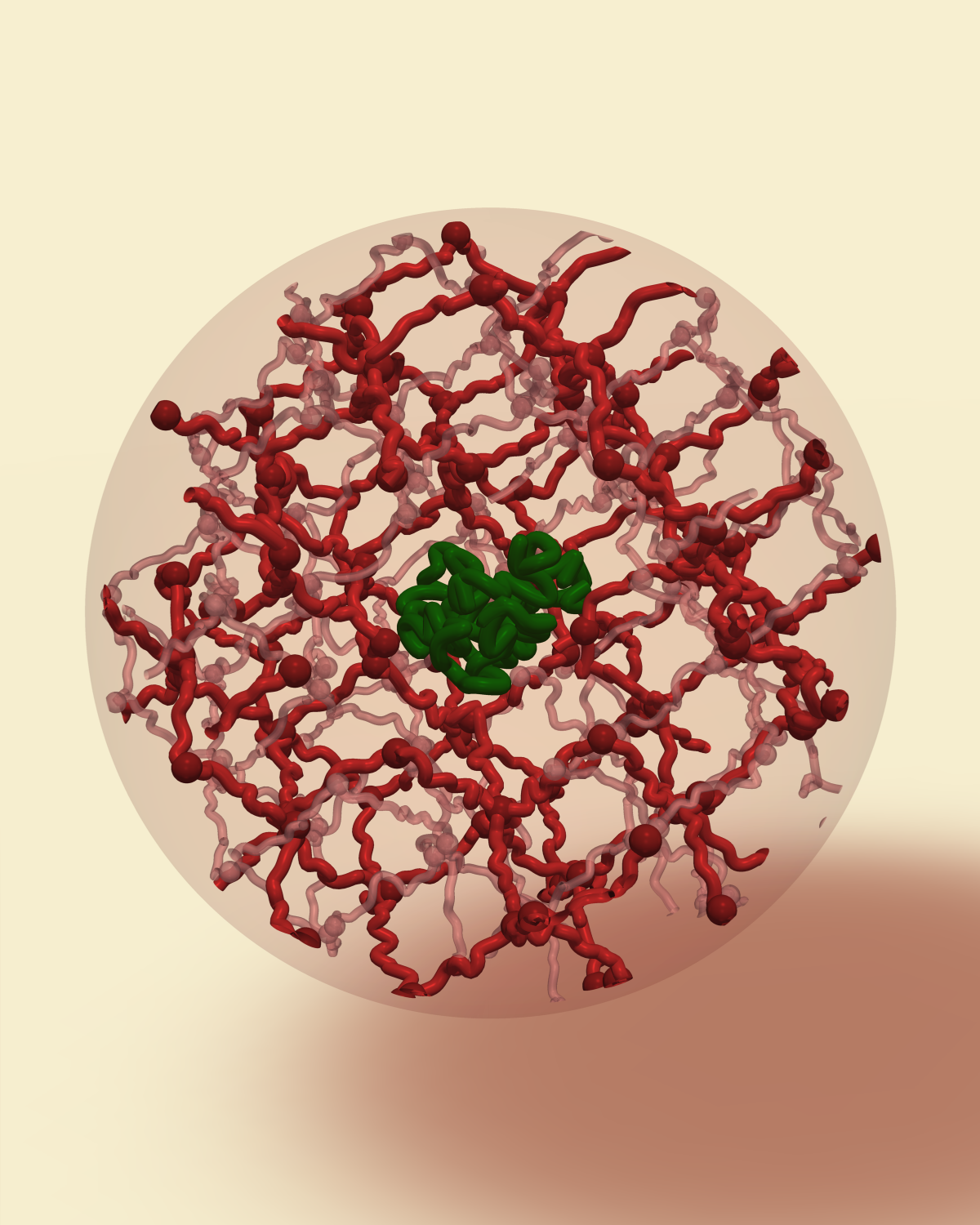}
\end{center}
\caption{Schematic depiction of a hydrogel (red cross-linked network) in osmotic equilibrium with a reservoir 
solution containing a free polymer coil (green) and solvent (yellow). The gel is fully permeable to solvent 
and selectively permeable to free polymer coils, whose degree of penetration depends on coil size 
relative to gel pore size.}
\label{fig:gel-reservoir}
\end{figure}

The model system consists of a hydrogel dispersed in a polymer solution (Fig.~\ref{fig:gel-reservoir}). 
The gel is fully permeable to solvent, but only partially permeable to free polymer coils. The degree 
of permeation depends on the relative sizes of coils and gel pores. The system is modeled in the 
semi-grand canonical ensemble, with fixed temperature and hydrogel polymer content and exchange of 
solvent and free polymer between the gel and a reservoir of free polymer solution. In thermodynamic
equilibrium, the gel adjusts its swollen volume and internal composition such that the 
chemical potentials of both solvent and free polymer coils are equal inside and outside of the gel.

We model a hydrogel as a network of cross-linked polymer chains, capable of swelling and deswelling 
between collapsed (dry) and fully swollen states. The network comprises polymer monomers linked 
together to form chains, which are cross-linked by a uniform distribution of cross-linkers. 
In the case of a spherical microgel, the dry state has radius $R_\mathrm{dry}$ and the swollen state 
radius $R_\mathrm{swollen}$. In a good solvent, in which polymer chains mutually repel, each chain's
conformation is described by a self-avoiding random walk (SAW)~\cite{deGennes1979,rubinstein2003polymer,
Doi2013,Dhar1978}. A statistical segment includes a fixed number of monomers and has a length equal to 
the Kuhn length $l$, which characterizes the chain stiffness. For example, a PNIPAM hydrogel in water, 
with monomer radius $r_\mathrm{mon}\simeq 0.3$ nm, has
\begin{equation}
  l \simeq 2\,r_\mathrm{mon} = 0.6\;\mathrm{nm}.
  \label{eq:kuhn}
\end{equation}
An uncrowded SAW chain of $M$ segments has a root-mean-square (rms) radius of gyration
\begin{equation}
  R_g^0 = CM^{\nu}l,
  \label{eq:Rg-chain}
\end{equation}
with Flory scaling exponent $\nu=0.588$ and amplitude $C=0.44108$~\cite{sciutto1996}.
The free polymer coils are similarly modeled as SAWs.

\subsection{Flory-Rehner Theory of Hydrogels}
\label{sec:binary}

\begin{figure}[h]
\begin{center}
\includegraphics[width=0.9\columnwidth,angle=0]{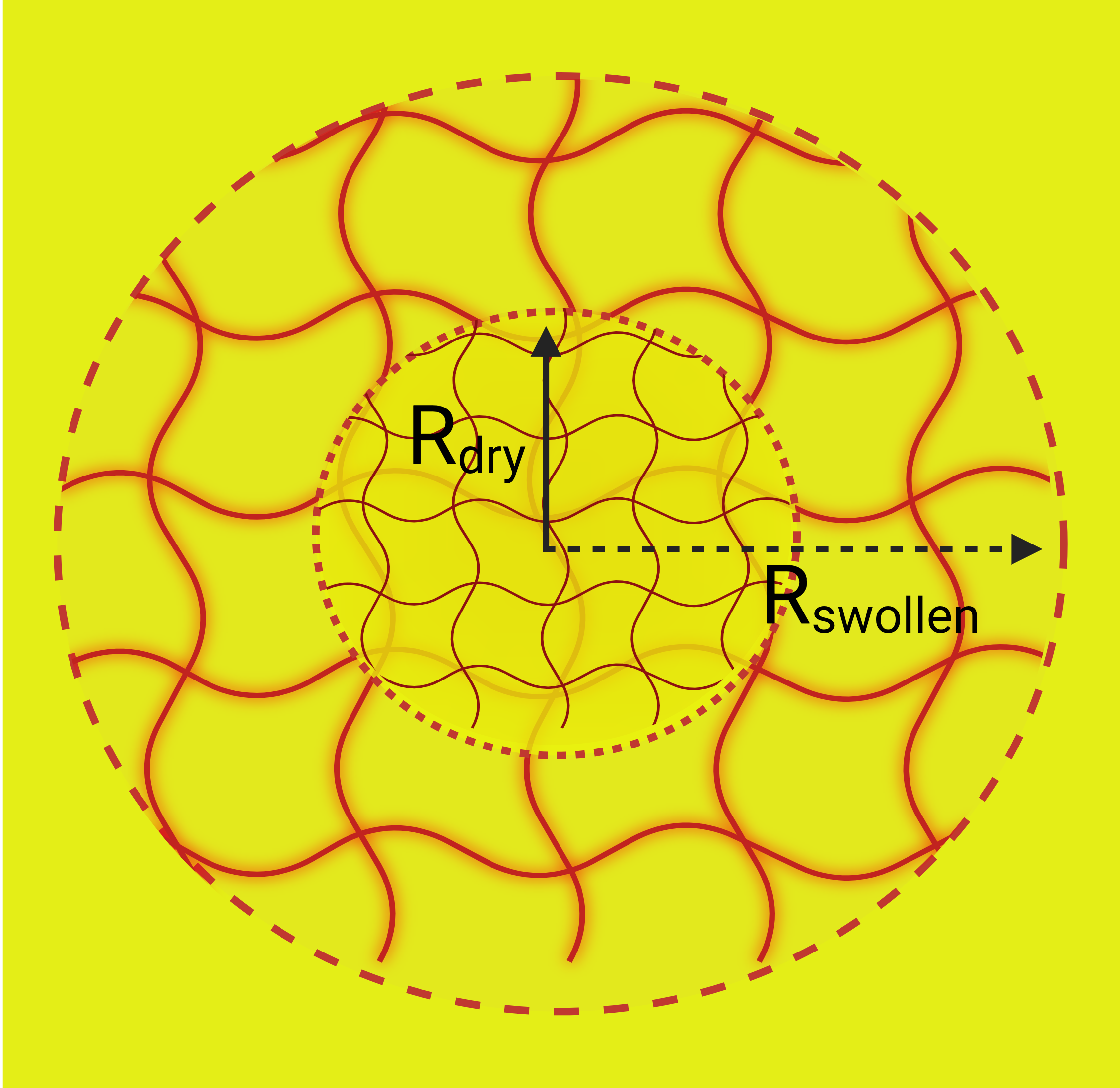}
\end{center}
\caption{Schematic illustration of a microgel (red cross-linked network) swelling between dry state 
of radius $R_{\rm dry}$ and swollen state of radius $R_{\rm swollen}$ in a solvent (yellow).
Created with BioRender.com/ugkahyw.}
\label{SingleMicrogelSwelling}
\end{figure}

Before introducing the free polymer coils, we recall the Flory-Rehner theory of a hydrogel, 
composed of cross-linked polymer chains (species~$p$), swollen by absorbing a solvent (species~$s$) 
(Fig.~\ref{SingleMicrogelSwelling}). The theory is based on the Flory-Huggins lattice model
of a polymer solution, in which solvent and polymer monomers are confined to sites on a lattice. 
Within this binary mixture model, the Helmholtz free energy of the hydrogel, comprising 
mixing entropy, polymer-solvent interaction energy, and network conformational entropy,
may be expressed as 
\begin{equation}
  \beta F_\mathrm{gel}
  = N_s \ln\!\left(\frac{v_p}{v_s}\,\phi_s\right)
  + \chi_{ps}\,N_p\,\phi_s
  + \frac{3N_{\rm ch}}{2}
    \bigl(\alpha^2 - 1 - \ln\alpha\bigr),
  \label{eq:F-binary}
\end{equation}
where $\beta\equiv 1/(k_BT)$, $N_s$ and $N_p$ are the respective numbers of solvent molecules and 
polymer monomers of the gel network, with molecular volumes $v_s$ and $v_p$, $\chi_{ps}$ is the 
Flory-Huggins interaction parameter, and $N_{\rm ch}$ is the number of chains in the network. 
In the case of tetravalent cross-linkers, the cross-link fraction is defined as 
$x_{\rm link}\equiv 0.5N_{\rm ch}/N_p$.
For a gel of swollen volume $V$, the solvent and polymer monomer volume fractions are 
$\phi_s = N_s v_s / V$ and $\phi_p = N_p v_p / V$. As the gel swells or deswells, $V$ changes 
and the total number of lattice sites occupied by the gel increases or decreases accordingly. 
Full occupancy of the lattice (no vacancies) requires $\phi_s + \phi_p = 1$, consistent with 
incompressibility.  Assuming affine deformation of the polymer network, the swelling ratio, 
which measures the linear expansion of the hydrogel relative to its dry state, is defined by
\begin{equation}
\alpha = \left(\frac{V}{V_{\rm dry}}\right)^{1/3},
\label{eq:alpha}
\end{equation}
where $V_{\rm dry}$ is the dry volume of the gel. For a spherical microgel, 
$\alpha=R_\mathrm{swollen}/R_\mathrm{dry}$ and $V_{\rm dry}=(4\pi/3)R_\mathrm{dry}^3$.

Next, we extend the binary mixture model of a hydrogel to a ternary mixture by introducing free polymer coils 
as a third species. Within the Flory-Huggins lattice framework, the interior of the hydrogel now includes,
in addition to solvent molecules and a cross-linked network of polymer monomers, $N_c$ free polymer coils, 
each comprising $M$ statistical segments (molecular weight) of volume $v_f$ with monomer volume fraction 
$\phi_f = N_c M v_f / V$. The corresponding coil volume fraction is then
$\phi_c = N_c v_c / V$, where $v_c$ is the coil volume.
For simplicity, we assume that the molecular volumes of the solvent, network polymer, and free polymer 
monomers are all equal ($v_s=v_p=v_f$), ensuring that each particle occupies a single lattice site. 
The incompressibility condition now becomes
\begin{equation}
  \phi_s + \phi_p + \phi_f = 1.
  \label{eq:incompress}
\end{equation}

\subsection{Free Polymer Coils as Fluctuating Ellipsoids}
\label{sec:ellipsoid}

Similar to the polymers that make up the hydrogel, the free polymer coils are modeled as 
self-avoiding random walks~\cite{deGennes1979,rubinstein2003polymer,Doi2013,Dhar1978}. 
In order to apply free-volume theory, we represent the conformation 
of a coil by an ellipsoid~\cite{kuhn1934shape} whose fluctuating size and shape are tied to random walk statistics.
The principal radii of the ellipsoid are related to eigenvalues of the gyration tensor of the 
coil in a frame of reference that rotates with the principal axes~\cite{Theodorou1985,rudnick1987shapes,
lim-denton-JCP2014,lim-denton-JCP2016,lim-denton-SM2016,davis-denton-JCP2018}.
For a coil of $M$ segments at positions $\mathbf{r}_i$, the gyration tensor
\begin{equation}
  \overleftrightarrow{\mathcal{R}}
  = \frac{1}{M}\sum_{i=1}^M
    (\mathbf{r}_i - \mathbf{r}_{cm})
    (\mathbf{r}_i - \mathbf{r}_{cm})^{\mathrm{T}}
  \label{eq:gyration}
\end{equation}
characterizes the spatial distribution of segments around the center of mass position $\mathbf{r}_{cm}$.

Diagonalizing $\overleftrightarrow{\mathcal{R}}$ yields eigenvalues $\Lambda_1\geq\Lambda_2\geq\Lambda_3$, 
which are proportional to the squared principal radii of an equivalent ellipsoid. 
The trace of the gyration tensor gives the instantaneous squared radius of gyration 
for a particular conformation:
\begin{equation}
  R_p^2 = \mathrm{Tr}\bigl(\overleftrightarrow{\mathcal{R}}\bigr)
        = \Lambda_1 + \Lambda_2 + \Lambda_3.
  \label{eq:Rg2}
\end{equation}
The moment of inertia tensor relative to the center of mass is related to the gyration tensor by
\begin{equation}
  \overleftrightarrow{\mathcal{I}}
  = R_p^2\,\mathbf{1} - \overleftrightarrow{\mathcal{R}},
  \label{eq:inertia}
\end{equation}
where $\mathbf{1}$ denotes the identity tensor.

Monte Carlo simulations of polymers on a lattice~\cite{sciutto1996} have determined the 
average eigenvalue probability distribution of a SAW to be accurately described by 
\begin{equation}
P_0(\lambda_1,\lambda_2,\lambda_3) = \prod_{i=1}^3 P_{i0}(\lambda_i),
\label{P0}
\end{equation}
where the three factors are given by
\begin{equation}
P_{i0}(\lambda_i)=a_i \lambda_i^{b_i} \exp(-c_i\lambda_i).
\label{Pi0-scaled}
\end{equation}
Here $\lambda_i\equiv\Lambda_i/(M^{\nu}l)^2$ are scaled eigenvalues, ordered according to
$\lambda_1 > \lambda_2 > \lambda_3$, and $a_i$, $b_i$, and $c_i$ are fit parameters
tabulated in Table~\ref{table1}.
The factorized form of Eq.~(\ref{P0}) assumes that the eigenvalues are all independent, which
proves accurate for sufficiently long polymers, excepting improbable conformations with 
extension primarily in one direction. 
\begin{table}
\caption{Parameters for shape distribution in Eq.~(\ref{Pi0-scaled}).}
\centering
\begin{tabular}{|c|c|c|c|}
\hline
eigenvalue $i$ & $a_i$ & $b_i$ & $c_i$ \\
\hline
\hline
1& 1.18479$\times 10^4$ & 2.35505 & 22.3563 \\
\hline
2& 1.11669$\times 10^9$ & 3.71698 & 148.715 \\
\hline
3& 1.06899$\times 10^{14}$ & 4.84822 & 543.619 \\
\hline
\end{tabular}
\label{table1}
\end{table}

The conformational statistics of a (possibly confined) polymer coil are encoded in the 
joint probability density, or shape distribution $P(\lambda_1,\lambda_2,\lambda_3)$~\cite{Arkin2013}.
Two important shape descriptors are the rms radius of gyration,
\begin{equation}
  R_g = \frac{R_g^0}{C}\sqrt{\langle\lambda_1+\lambda_2+\lambda_3\rangle},
  \label{eq:Rg-avg}
\end{equation}
where $R_g^0$ is the rms radius of gyration of an unconfined coil [Eq.~\eqref{eq:Rg-chain}], 
and the asphericity,
\begin{equation}
  A = 1 - 3\frac{\langle\lambda_1\lambda_2 + \lambda_1\lambda_3 + \lambda_2\lambda_3\rangle}
       {\langle(\lambda_1+\lambda_2+\lambda_3)^2\rangle},
  \label{eq:asphericity}
\end{equation}
where angular brackets denote ensemble averages over the shape distribution.
The asphericity satisfies $0\leq A\leq 1$, vanishing for spherically symmetric conformations 
and approaching unity for highly elongated, rod-like shapes. Strong confinement skews
the eigenvalue distribution toward smaller, more isotropic shapes, thereby reducing both 
$R_g$ and $A$, a signature of conformational compaction analogous to protein folding.
From the statistics of a SAW, we estimate the free polymer coil volume in the reservoir 
as~\cite{davis-denton-JCP2018}
\begin{equation}
  v_c \simeq 1.693\,(R_g^0)^3.
  \label{eq:vcoil}
\end{equation}

\subsection{Free-Volume Theory of Hydrogel Network}
\label{sec:confinement}
\begin{figure}[h]
\includegraphics[width=\columnwidth,angle=0]{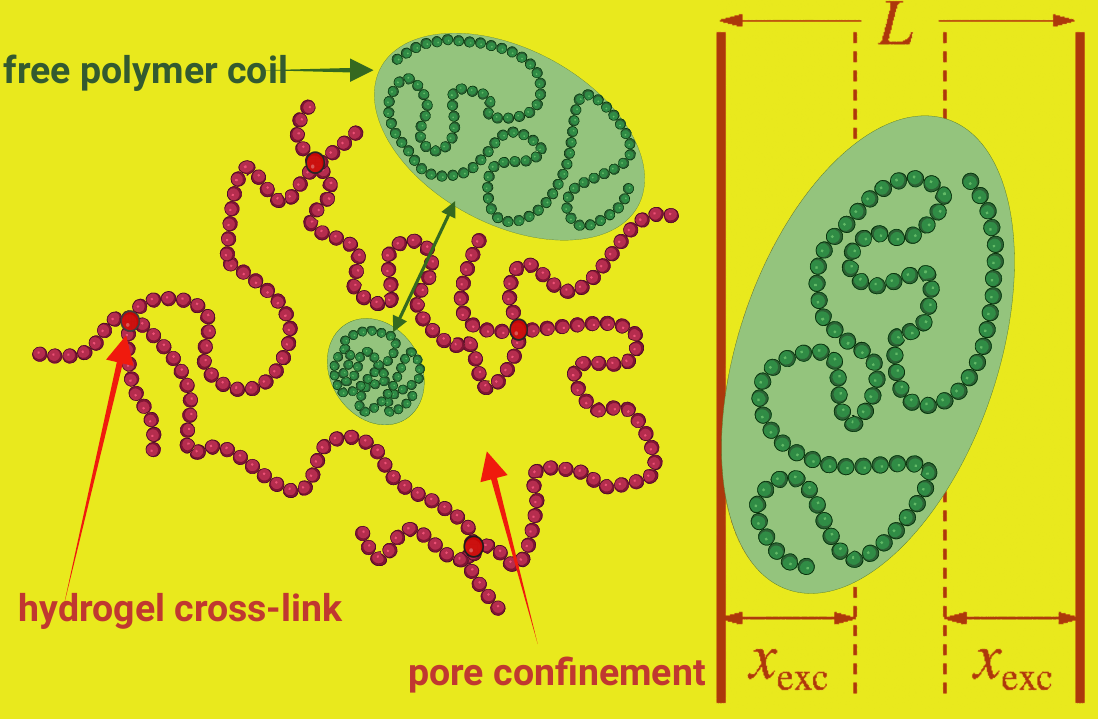}
\caption{Left: Microscopic view of confinement within hydrogel cross-links. 
Right: Confinement between cross-links is approximated by a slit pore: 
parallel hard walls separated by distance $L$, proportional to gel swelling ratio.
Created with BioRender.com/gfwar3o.}
\label{Confinement}
\end{figure}
 
Mapping the conformation of a free polymer coil onto a solid body of well-defined geometry
enables the application of concepts from free-volume theory~\cite{lekkerkerker1992}.
To describe the confinement of free polymer coils within a swollen hydrogel, specifically
within the space (pores) between neighboring cross-links, we approximate the confining geometry 
by a slit pore, defined by two parallel, hard walls separated by a distance $L$ (Fig.~\ref{Confinement}).
The pore size (slit width) scales in proportion to the swelling ratio: $L=L_0\alpha$, where 
$L_0$ is a reference pore size. As the hydrogel swells, $L$ increases, enlarging 
the available free volume. Although somewhat idealized, the slit pore model captures the main 
physical effects of confinement while greatly simplifying numerical calculations. 

When confined within a slit pore of width $L$, an ellipsoid, modeling a free polymer coil, whose shape 
is characterized by gyration tensor eigenvalues, $\lambda\equiv\{\lambda_1,\lambda_2,\lambda_3\}$, 
and whose orientation $\Omega$ is specified by polar and azimuthal angles, 
has a free-volume fraction defined as
\begin{equation}
\kappa(\lambda, \Omega, L, M) \equiv 1 - \frac{V_\mathrm{exc}(\lambda, \Omega, L, M)}{V},
\label{eq:kappa}
\end{equation}
where $V_\mathrm{exc}(\lambda, \Omega, L, M)$ is the volume excluded to the center of mass
of the coil (ellipsoid).
Averaging $\kappa(\lambda, \Omega, L)$ over all polymer shapes and orientations yields
an {\it effective} free-volume fraction:
\begin{equation}
\kappa_{\rm eff}(L, M) = \int d\lambda\int d\Omega\, \kappa(\lambda, \Omega, L, M) P_0(\lambda),
\label{eq:kappaeff}
\end{equation}
where the angular integrals cover the full $4\pi$ solid angle and the eigenvalue integrals are
abbreviated as
\begin{equation}
\int d\lambda \equiv\int_0^{\infty}\!d\lambda_1\!\int_0^{\lambda_1}\!d\lambda_2
\!\int_0^{\lambda_2}\!d\lambda_3,
\label{eq:lambda}
\end{equation}
respecting the ordering, $\lambda_1>\lambda_2>\lambda_3$. As illustrated in Fig.~\ref{Confinement},
the effective free-volume fraction can also be expressed as $1-2x_{\rm exc}/L$, where $x_{\rm exc}$ 
is the mean width of the depletion layer of the polymer adjacent to the wall. The latter quantity, 
which equals the integrated mean curvature of the ellipsoid, is computed from the shortest distance 
between the ellipsoid center and a wall (i.e., half-width of ellipsoid)~\cite{denton-davis-JCP2021} .
We note in passing that the property defined in Eq.~\eqref{eq:kappaeff} is equivalent to the 
partition coefficient $\kappa$ introduced elsewhere to describe how confinement restricts 
accessible configurational space~\cite{Ping2004,Zhou2008,Takarai2026}.

Free-volume theory predicts the probability distribution for the shape of a confined polymer 
coil~\cite{lim-denton-JCP2014,lim-denton-JCP2016,lim-denton-SM2016,davis-denton-JCP2018}:
\begin{equation}
P(\lambda, \Omega, L, M) = \frac{\kappa(\lambda, \Omega, L, M)}{\kappa_{\rm eff}(L, M)} P_0(\lambda).
\label{P}
\end{equation}
With respect to this probability distribution, the ensemble average of any quantity $Q$ is
\begin{equation}
\la Q\ra = \int d\lambda\int d\Omega\, Q P(\lambda, \Omega, L, M).
\label{Qaverage}
\end{equation}
In particular, the radius of gyration and asphericity of a confined polymer coil can be
computed from Eqs.~\eqref{eq:Rg-avg}, \eqref{eq:asphericity}, \eqref{P}, and \eqref{Qaverage}.

\section{Free Energy}
\label{sec:free-energy}

Combining the Flory-Rehner theory of polymer networks with the free-volume theory of confined 
polymers, we now collect the three fundamental contributions to the Helmholtz free energy of a 
hydrogel in equilibrium with a reservoir solution of free polymer of monomer
volume fraction $\phi_f^0$.
The two mobile species -- solvent molecules and free polymer coils -- have translational entropies 
that together make a contribution
\begin{equation}
  \beta F_\mathrm{ent}
  = N_s\ln\phi_s
  + N_c\ln\left[\frac{\rho_c(L, M)v_p}{\kappa_\mathrm{eff}(L, M)}\right],
  \label{eq:F-ent}
\end{equation}
where $\rho_c\equiv N_c/V$ is the number density of coils in the total volume.
The first term on the right side is the standard lattice mixing free energy of the solvent. 
The second term is the translational entropic free energy of the coils,
assumed sufficiently dilute as to behave like an ideal-gas, but confined to the effective 
(lower) free volume $\kappa_\mathrm{eff}(L, M)V$ of the hydrogel pores of size $L$.
To be clear, $\rho_c$ represents the global number density and $\rho_c/\kappa_\mathrm{eff}$
the local number density of free polymer coils. The former is always lower than the latter
because of excluded volume due to confinement by the hydrogel network.

Within the mean-field (Flory-Huggins) approximation, the energetic contribution to the 
mixing free energy is
\begin{equation}
  \beta E_\mathrm{mix}
  = \chi_{ps}\,N_p\,\phi_s
  + \chi_{pf}\,N_p\,\phi_f
  + \chi_{sf}\,N_s\,\phi_f,
  \label{eq:E-mix}
\end{equation}
where the three terms account, respectively, for polymer-solvent, polymer-coil, and solvent-coil 
nearest-neighbor interactions and $\chi_{ij}$ denote the standard Flory-Huggins 
pairwise interaction parameters. 

Adopting an affine swelling model of the gel, the elastic free energy of the 
cross-linked network is~\cite{FloryRehner1943a,FloryRehner1943b,Erman1986}
\begin{equation}
  \beta F_\mathrm{el}
  = \frac{N_{\rm ch}}{2}\bigl(\alpha_x^2 + \alpha_y^2 + \alpha_z^2 - 3 - \ln\alpha_x\alpha_y\alpha_z\bigr).
  \label{eq:F-el-aniso}
\end{equation}
For isotropic swelling, $\alpha_x = \alpha_y = \alpha_z \equiv \alpha$, this contribution reduces to
\begin{equation}
  \beta F_\mathrm{el} = \frac{3}{2}N_{\rm ch}\bigl(\alpha^2 - 1 - \ln\alpha\bigr).
  \label{eq:F-el}
\end{equation}

Summing Eqs.~\eqref{eq:F-ent}-\eqref{eq:F-el} yields the total free energy of the ternary 
gel-solvent-coil mixture:
\begin{align}
  \beta F
  &= N_s\ln\phi_s + N_c\ln\left[\frac{\rho_c(L, M)v_p}{\kappa_\mathrm{eff}(L, M)}\right]
   + \chi_{ps}\,N_p\,\phi_s
  \notag\\[1ex]
  &
   + \chi_{pf}\,N_p\,\phi_f
   + \chi_{sf}\,N_s\,\phi_f
   + \frac{3}{2}N_{\rm ch}\left(\alpha^2 - 1 - \ln\alpha\right).
  \label{eq:F-total}
\end{align}
Note that the confinement geometry enters only via $\kappa_\mathrm{eff}(L, M)$. 
Minimizing $F(\phi_s,\phi_f,\alpha)$ with respect to $\phi_s$ and $\phi_f$, 
subject to incompressibility [Eq.~\eqref{eq:incompress}], is equivalent to equating 
chemical potentials inside and outside the hydrogel and yields the equilibrium 
swelling ratio and internal composition of the gel.

\section{Chemical Potentials}
\label{sec:chempot}

Chemical equilibrium requires equality of chemical potentials of free polymer coils 
inside and outside the gel:
\begin{equation}
\mu_f^{\mathrm{in}}=\mu_f^0.
\label{eq:muf-equality}
\end{equation}
These chemical potentials result from the partial derivative of $F$ with respect to $N_c$ 
(holding fixed $N_s$):
\begin{equation}
  \beta\mu_f^{\mathrm{in}}
  = \ln\left[\frac{\rho_c(L, M)v_p}{\kappa_\mathrm{eff}(L, M)}\right],
  \qquad
  \beta\mu_f^0
  = \ln\bigl(\rho_c^0\,v_p\bigr),
  \label{eq:mu-fp}
\end{equation}
where $\rho_c(L, M)$ and $\rho_c^0$ are the coil densities inside the gel and in the reservoir, respectively. 
It follows immediately from Eq.~\eqref{eq:muf-equality} that
\begin{equation}
\rho_c(L, M) = \rho_c^0\kappa_\mathrm{eff}(L, M),
\qquad 
\phi_f(L) = \phi_f^0\kappa_\mathrm{eff}(L, M).
\label{eq:rho-phi-fp}
\end{equation}

Chemical equilibrium further requires equality of solvent chemical potentials 
inside and outside the gel:
\begin{equation}
\mu_s^{\mathrm{in}}(L, \phi_f, \phi_f^0, M) = \mu_s^0(\phi_f^0, M).
\label{eq:mus-equality}
\end{equation}
The solvent chemical potential inside the gel is obtained by taking the partial derivative of $F$ 
[Eq.~\eqref{eq:F-total}] with respect to $N_s$ (now holding fixed $N_c$):
\begin{align}
  \beta\mu_s^{\mathrm{in}}
  &= \ln\phi_s + 1 - \phi_s
  - \frac{\phi_f^0}{M}
    \left(
      \kappa_\mathrm{eff} + \frac{\alpha}{3}L_0\frac{\partial\kappa_\mathrm{eff}}{\partial L}
    \right)
  \notag\\[1ex]
  &
  + \chi_{ps}\,\phi_p\,(1-\phi_s)
  - \chi_{pf}\,\phi_p\,\phi_f
  + \chi_{sf}\,\phi_f\,(1-\phi_s)
  \notag\\[1ex]
  &
  + x_{\rm link}\,(2\alpha^2-1)\,\phi_s, 
  \label{eq:mu-s-in}
\end{align}
where we have used Eq.~(\ref{eq:rho-phi-fp}).
Outside the gel, in the reservoir, where the free volume equals the total volume 
($\kappa_\mathrm{eff}=1$), only the ideal-mixing and solvent-coil interaction terms survive:
\begin{equation}
  \beta\mu_s^0
  = \ln\phi_s^0 + 1 - \phi_s^0
  - \frac{\phi_f^0}{M}
  + \chi_{sf}\,\phi_f^0\,(1 - \phi_s^0),
  \label{eq:mu-s-res}
\end{equation}
with $\phi_f^0 = 1 - \phi_s^0$.

For specified values of free polymer concentration and molecular weight ($\phi_f^0$ and $M$), 
simultaneous solution of Eqs.~\eqref{eq:rho-phi-fp} and \eqref{eq:mus-equality} for the two unknowns 
($L$ and $\phi_f$) yields both the equilibrium state of the swollen hydrogel and the internal 
composition of the gel, both of which depend on a single geometric factor, $\kappa_\mathrm{eff}(L, M)$.

\section{Computational Methods}
\label{sec:computation}

To compute the effective free-volume fraction $\kappa_\mathrm{eff}(L, M)$ as a function of pore size $L$, 
we numerically evaluated the five-dimensional integral of Eq.~\eqref{eq:kappaeff}, using Gaussian 
quadrature. We then parametrized the data by fitting to a function of the reduced wall separation 
$x\equiv L/R_g^0$, taking the radius of gyration of an unconfined (reservoir) SAW chain $R_g^0$ 
[Eq.~\eqref{eq:Rg-chain}] as a fixed reference length. An analytical fit function, 
accurate for $x>1$, is
\begin{equation}
  \kappa_\mathrm{eff}(x) = 1 - \frac{1.8}{x} + 0.8e^{-2(x-1)}.
  \label{eq:alpha-fit}
\end{equation}

Next, we computed the equilibrium swelling ratio $\alpha$ and internal composition of the hydrogel $\phi_f$ 
by solving the set of nonlinear equations derived in Sec.~\ref{sec:chempot} [Eqs.~\eqref{eq:mus-equality} 
and \eqref{eq:rho-phi-fp}] over a range of reservoir free polymer concentration $\phi_f^0$ and for 
specific choices of coil molecular weight. For this purpose, we implemented the Levenberg-Marquardt 
root-finding algorithm in Python via the SciPy \texttt{fsolve} function. As an initial estimate, 
we used the pure-solvent equilibrium value of $\phi_p^0$ and updated $\phi_p$ sequentially 
along the concentration sweep. 

From $\alpha$ we obtained the equilibrium pore size $L=L_0\alpha$, 
where $L_0=(4\pi/3x_\mathrm{link})^{1/3}\,r_\mathrm{mon}$
is the reference pore size in the dry state. 
Because $L$ depends on $\alpha$, which itself depends 
on $\kappa_\mathrm{eff}(L/R_g^0)$ through the chemical-potential balance, the confinement geometry 
and thermodynamic state of the gel are coupled and must be determined simultaneously.  
For each molecular weight, the unconfined radius of gyration [Eq.~\eqref{eq:Rg-chain}] sets the 
confinement length scale and the reduced pore size $x$ determines the position on the universal 
$\kappa_\mathrm{eff}(x)$ curve [Eq.~\eqref{eq:alpha-fit}].

\section{Results and Discussion}
\label{sec:results}

We now present predictions for the equilibrium swelling response and internal composition of a hydrogel 
immersed in a reservoir solution of free polymer coils. Unless otherwise stated, the gel cross-link 
fraction is fixed at $x_\mathrm{link}=5\times 10^{-4}$ and all Flory-Huggins interaction parameters 
are set to zero ($\chi_{ps} = \chi_{sf} = \chi_{pf} = 0$), corresponding to athermal (purely entropic) 
conditions. We determined the equilibrium swelling ratio and internal free polymer concentration by 
numerically finding the roots of the chemical-potential balance conditions [Eqs.~\eqref{eq:mus-equality} 
and \eqref{eq:mu-fp}].
Most results presented below are for three representative coil molecular weights: $M = 100$, $300$, and $500$.

\subsection{Effective Free-Volume Fraction}
\label{sec:results-Keff}
\begin{figure}[h]
\includegraphics[width=\columnwidth,angle=0]{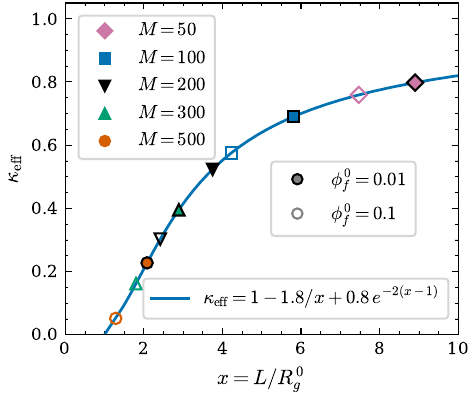}
\caption{Effective free-volume fraction $\kappa_\mathrm{eff}$ vs.~reduced pore size $x \equiv L/R_g^0$
from numerical evaluation of Eq.~\eqref{eq:kappaeff} for SAW chains, modeled as fluctuating ellipsoids, 
confined between hard walls. The solid curve is the fit function of Eq.~\eqref{eq:alpha-fit},
which is universal for all molecular weights. Filled and open symbols mark the equilibrium values 
at free polymer reservoir concentrations $\phi_f^0 = 0.01$ and $0.1$, respectively, for molecular weights 
$M$ in the range 50-500 (right to left). Increasing $\phi_f^0$ shifts each point leftward along the curve, 
reflecting gel deswelling and tighter confinement.  
}
\label{fig:kappa-eff}
\end{figure}

Figure~\ref{fig:kappa-eff} shows the effective free-volume fraction $\kappa_\mathrm{eff}(L)$ of a 
SAW chain, modeled as a fluctuating ellipsoid, confined to a slit pore of reduced pore size $x \equiv L/R_g^0$.
The curve is a plot of the function [Eq.~\eqref{eq:alpha-fit}] that fits data from numerical evaluation of
Eq.~\eqref{eq:kappaeff} for SAW chains modeled as fluctuating ellipsoids confined between hard walls.
Because the shape statistics of SAW chains are self-similar when scaled by
$R_g^0$, the function $\kappa_\mathrm{eff}(x)$ is universal:
all molecular weights collapse onto the same curve.

Three confinement regimes are evident. For $x>5$ (weak confinement), $\kappa_\mathrm{eff}>0.5$ 
and the majority of chain conformations are compatible with the geometric restriction imposed by the walls.  
In the moderate-confinement regime ($1<x<5$), $\kappa_\mathrm{eff}$ drops steeply, as a 
progressively smaller fraction of conformations fit between the walls. For $x<1$ (strong confinement), 
$\kappa_\mathrm{eff} \to 0$ and the polymer chains are effectively excluded from the pore.

The filled and open symbols in Fig.~\ref{fig:kappa-eff} mark the equilibrium  points for different 
free polymer coil molecular weights at reservoir concentrations $\phi_f^0 = 0.01$ and $0.1$, respectively.
The equilibrium pore size $L$ and the unconfined radius of gyration $R_g^0(M)$ together determine 
where each point $(M, \phi_f^0)$ sits on the universal curve. 
The key observation is that different molecular weights access different regions of this curve. 
Increasing $\phi_f^0$ from $0.01$ to $0.1$ shifts each point leftward along the curve, 
as the gel deswells, the pores shrink, and the confinement tightens.

The universal curve in Fig.~\ref{fig:kappa-eff} is the SAW version of the partition function 
computed by Zhou and Dill~\cite{ZhouDill2001} for a Gaussian chain between two parallel walls 
[their Eq.~(3)]. Since SAW chains are more extended than Gaussian chains (Flory exponent $\nu = 0.588$ 
vs.~$0.5$), confinement effects set in at larger values of $L/R_g^0$.  
Unlike the Zhou-Dill calculation, in which the wall separation is fixed, our pore size responds 
self-consistently to the free polymer loading through the swelling equilibrium, coupling the 
confinement geometry to the thermodynamic state of the gel.  

\begin{figure}[h]
\includegraphics[width=\columnwidth,height=8cm]{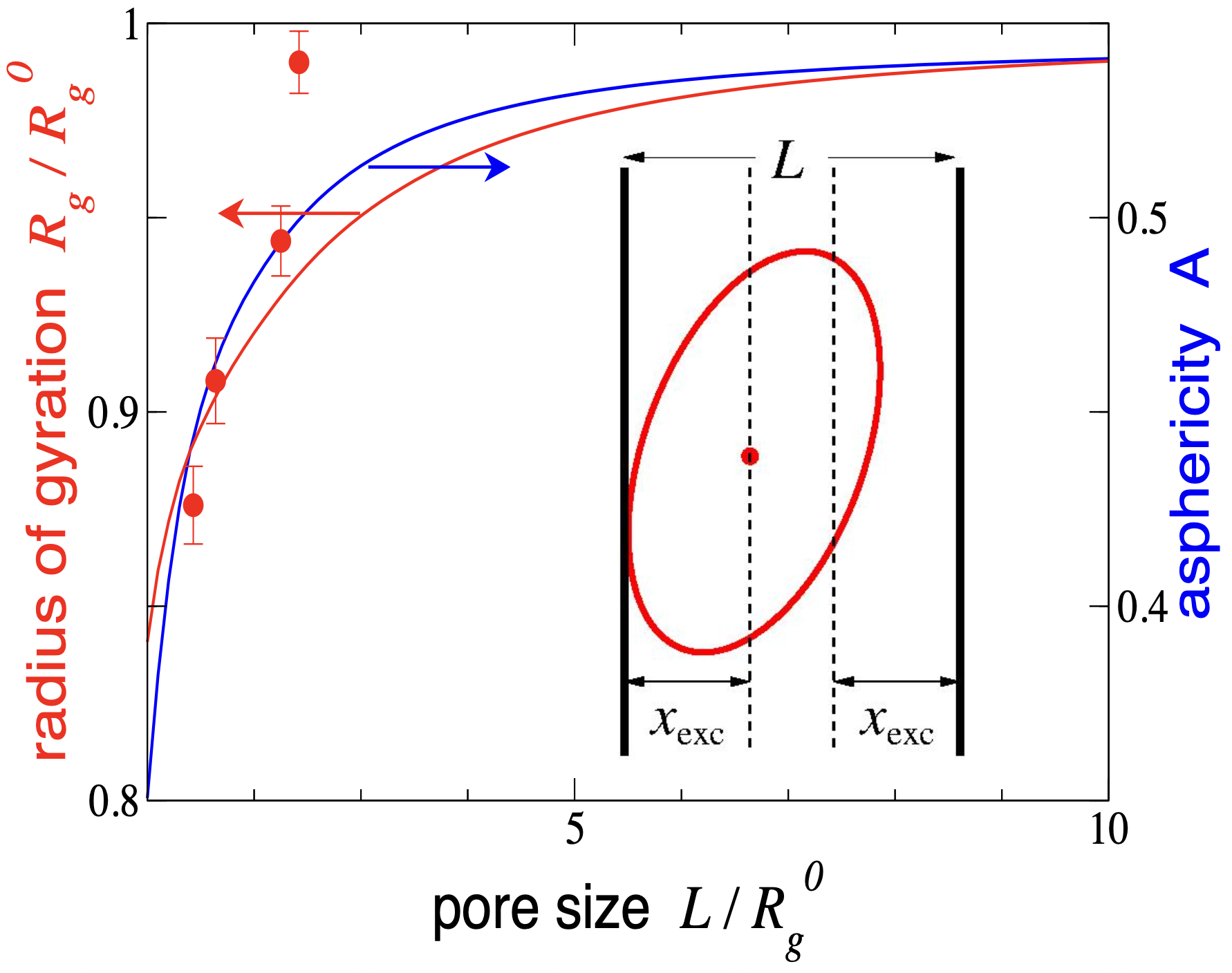}
\caption{Predictions for normalized radius of gyration $R_g/R_g^0$ (red curve, left axis) 
and asphericity $A$ (blue curve, right axis) of a confined SAW polymer coil as functions 
of reduced pore size $L/R_g^0$. Also shown are experimental SANS data for $R_g/R_g^0$ 
 \cite{briber1995,briber1997} (red circles).
Inset: schematic of a polymer coil (fluctuating ellipsoid) confined between hard walls 
separated by distance $L$, showing excluded-volume zones $x_\mathrm{exc}$ from which the 
polymer center of mass is sterically excluded.}
\label{fig:confinement}
\end{figure}

\subsection{Conformations of Confined Free Polymer}
\label{sec:results-aeff}
From the effective free-volume fraction, we computed the radius of gyration and asphericity 
of a confined free polymer coil via Eqs.~\eqref{eq:Rg-avg}, \eqref{eq:asphericity}, \eqref{P}, 
and \eqref{Qaverage}. Our results for $R_g/R_g^0$ and $A$ as functions of the reduced pore size 
$L/R_g^0$ are shown in Fig.~\ref{fig:confinement}. These two quantities characterize, respectively, 
the size and shape of the confined polymer. Their simultaneous variation with pore size 
provides a direct measure of confinement-induced conformational changes.

In the wide-pore (highly swollen) limit ($L/R_g^0 \gtrsim 10$), both $R_g$ and $A$ converge to 
their bulk values for a SAW coil: $R_g/R_g^0 \to 1$ and $A \to 0.54$, the polymer then being 
unperturbed by the confining walls. As the pore narrows below $L/R_g^0 \simeq 5$, 
confinement effects become increasingly evident. The radius of gyration decreases to 
$R_g/R_g^0 \simeq 0.8$ at the strongest confinements ($1<L/R_g^0\lesssim 2$), indicating
compression of the polymer coil to roughly 80\% of its unconfined size. Simultaneously, the
asphericity drops to $A \simeq 0.38$, reflecting a shift from the prolate, elongated conformations 
typical of unconfined SAW chains toward more compact, isotropic shapes.

Also shown in Fig.~\ref{fig:confinement} are data (red circles) for $R_g/R_g^0$ from small-angle 
neutron scattering (SANS) experiments of Briber et al.~\cite{briber1995,briber1997}, which probed 
the conformations of free linear chains of deuterated polystyrene trapped within a matrix comprising 
a cross-linked network of protonated polystyrene. For this comparison, we took $R_g^0=\sqrt{M/6}l$,
assuming that free chains have Gaussian statistics in the melt, and $L=\sqrt{N_c}l$, the mean
end-to-end displacement of network chains, where $N_c$ is the number of segments between cross-links.
Error bars are estimated from the reported uncertainties in $R_g$ and $R_g^0$.

Our predictions are evidently in qualitative agreement with the SANS data, although the
experimental $R_g$ approaches $R_g^0$ more rapidly with increasing $L$ than in our model. 
The shrinkage and compaction of confined free polymers predicted here is also somewhat analogous 
to the collapse of polymer chains in a porous medium of fixed obstacles observed in the 
Monte Carlo lattice simulations of Honeycutt and Thirumalai~\cite{honeycutt1989}, except that
the pores of our hydrogel medium are not static, but adapt their size to polymer loading.

The physical mechanism for compactification is illustrated in the inset of Fig.~\ref{fig:confinement}, 
which depicts a polymer coil (modeled as a fluctuating ellipsoid) confined between two hard walls 
separated by a distance $L$. The walls impose excluded-volume zones of width $x_\mathrm{exc}$ on either
side, from which the center of mass of the polymer is sterically excluded. As the wall separation narrows,
these exclusion zones occupy an increasingly large fraction of the total pore volume, precluding
extended conformations whose spatial extent exceeds the pore width. The surviving conformations are 
preferentially compact and isotropic, producing the observed reductions in both $R_g$ and $A$.

\subsection{Swelling Response to Free Polymer Crowding}
\label{sec:results-swelling}
\begin{figure}[h]
\includegraphics[width=\columnwidth,angle=0]{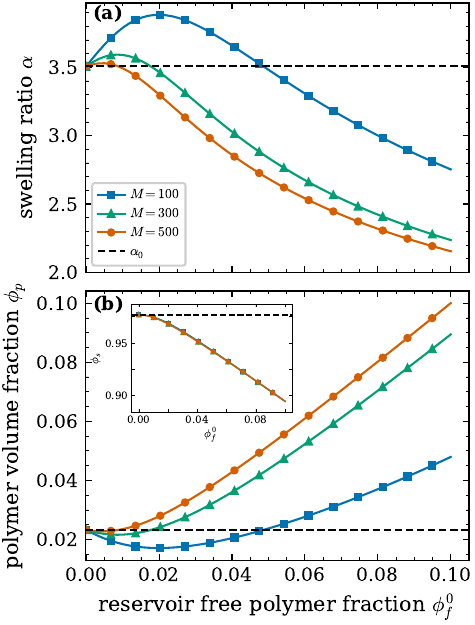}
\caption{Equilibrium swelling response of a hydrogel (cross-link fraction $x_\mathrm{link}=5\times 10^{-4}$) 
in a good-solvent solution of free polymer of reservoir monomer volume fraction $\phi_f^0$ and molecular weight 
$M=100$ (squares), $300$ (triangles), and $500$ (circles). (a) Swelling ratio $\alpha=\phi_p^{-1/3}$ 
and (b) network polymer monomer volume fraction $\phi_p$ vs.~$\phi_f^0$. Inset: Solvent volume fraction 
$\phi_s$ monotonically decreases, independent of $M$, as free polymer displaces solvent.
Horizontal dashed lines represent pure-solvent limit ($\phi_f^0=0$).
}
\label{fig:alpha-vs-phi}
\end{figure}

Figure~\ref{fig:alpha-vs-phi} shows results of our calculations for the equilibrium swelling ratio $\alpha$ 
and corresponding network polymer volume fraction $\phi_p$ as functions of reservoir free polymer loading 
(monomer volume fraction $\phi_f^0$) for three molecular weights, $M = 100$, $300$, and $500$. 
Interestingly, the response with varying $\phi_f^0$ is non-monotonic in all cases. With increasing 
free polymer loading, the gel initially swells as free polymer penetrates the interior, but then deswells 
at higher $\phi_f^0$.

This crossover from swelling to deswelling with increasing free polymer loading reflects a competition 
between two effects. At low reservoir concentrations, the osmotic pressure of absorbed coils drives 
additional solvent uptake and swells the gel. At higher reservoir concentrations, however, the external 
osmotic pressure overcomes the internal contribution, expelling solvent, and driving deswelling. 
As $\phi_f^0$ increases, the confinement-coupling contribution in the solvent chemical potential 
[$\kappa_\mathrm{eff}'$ term in Eq.~\eqref{eq:mu-s-in}] grows in magnitude and increasingly opposes 
swelling, because additional solvent entering the gel expands the pore at an increasing entropic cost.
This non-monotonic swelling response is qualitatively similar to the behavior observed experimentally 
by Vincent and coworkers~\cite{bradley-vincent2005,routh2006effect}, who reported that low concentrations 
of PEO chains permeate PNIPAM hydrogels and induce measurable swelling before external osmotic deswelling 
dominates at higher concentrations.

Furthermore, the height and concentration of the swelling peak both depend strongly on $M$. Evidently, 
the effect of free polymer is greatest for smaller coils, which can more readily penetrate the gel pores. 
For $M=100$, $\alpha$ rises well above the pure-solvent value $\alpha_0$, demonstrating significant 
network swelling, and peaks at $\phi_f^0 \simeq 0.02$ [Fig.~\ref{fig:alpha-vs-phi}(a)], while $\phi_p$ falls 
to a minimum at the same concentration [Fig.~\ref{fig:alpha-vs-phi}(b)]. Only beyond $\phi_f^0 \simeq 0.045$ 
do $\alpha$ and $\phi_p$ return to their baseline pure-solvent values. In contrast, for $M=500$, the 
swelling peak is much lower and occurs at lower free polymer loading ($\phi_f^0 \simeq 0.006$).
This behavior is consistent with the experimental observations of Saunders and Vincent~\cite{Saunders1996,Saunders1997a}, 
who found that the deswelling of polystyrene microgels increases with the molecular weight of added 
free polystyrene, due to larger chains being more strongly excluded from the gel interior.

The inset of Fig.~\ref{fig:alpha-vs-phi}(b) shows the solvent volume fraction $\phi_s$, which decreases 
for all three molecular weights as free polymer displaces solvent. The three curves nearly coincide, 
indicating that the internal solvent content is set primarily by $\phi_f^0$ and depends only weakly on $M$.

\subsection{Free Polymer Uptake and Partitioning}
\label{sec:results-partition}
\begin{figure}[h]
\includegraphics[width=\columnwidth,angle=0]{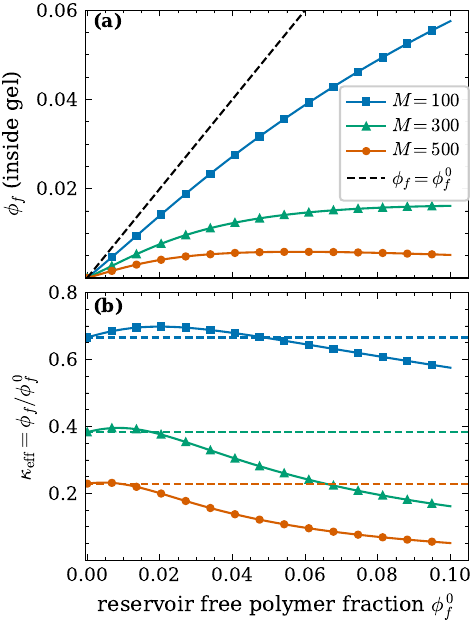}
\caption{Equilibrium internal composition of a hydrogel versus reservoir free polymer monomer volume fraction
$\phi_f^0$ for molecular weight $M=100$ (blue squares), $300$ (green triangles), and $500$ (orange circles).
(a)~Free-polymer monomer volume fraction $\phi_f$ inside the gel. Dashed line indicates $\phi_f=\phi_f^0$ 
(no exclusion), deviation below which reflects confinement-induced size exclusion, which increases
with $M$. For larger coils, the uptake saturates ($M=300$) or passes through a shallow maximum ($M=500$)
as the deswollen gel progressively tightens its pores.
(b)~Free-polymer monomer volume fraction $\phi_f$ inside gel relative to reservoir value $\phi_f^0$,
equal to effective free-volume fraction [Eq.~\eqref{eq:rho-phi-fp}].
Dashed lines indicate pure-solvent limit.
}
\label{fig:phi_fp-vs-phi}
\end{figure}

Because the pore size is proportional to the swelling ratio in our model,
the pore size mirrors the swelling response of Fig.~\ref{fig:alpha-vs-phi}(a). Starting from the
pure-solvent value $L=21.4$~nm, the $M=100$ pore expands to a maximum of 23.5 nm near
$\phi_f^0 \simeq 0.02$, then contracts, falling back through 21.4 nm near $\phi_f^0 \simeq 0.05$ 
and reaching 16.8 nm at $\phi_f^0 = 0.10$. For $M=300$ and $M=500$ the pore barely expands 
before contracting monotonically.

Figure~\ref{fig:phi_fp-vs-phi} displays the equilibrium free polymer uptake as a function of $\phi_f^0$.
The free polymer monomer volume fraction $\phi_f$ inside the gel [Fig.~\ref{fig:phi_fp-vs-phi}(a)] is 
in all cases lower than the reservoir concentration, because confinement excludes a fraction of chain 
conformations and reduces the effective free volume available to the coil centers of mass.
The degree of exclusion increases strongly with $M$. For $M=100$, 
the uptake is nearly linear with initial slope $\kappa_\mathrm{eff} \simeq 0.67$, curving
slightly sublinear at higher $\phi_f^0$ as the swelling saturates, and reaches $\phi_f \simeq 0.056$ at
$\phi_f^0 = 0.10$. For the larger coils the uptake is strongly suppressed and ceases to grow: the $M=300$ 
curve saturates near $\phi_f \simeq 0.016$, while the $M=500$ curve passes through a shallow maximum
($\phi_f \simeq 0.006$ near $\phi_f^0 \simeq 0.05$) and then slowly declines. At low concentration
($\phi_f^0=0.01$) the effective partition coefficients are $\kappa_\mathrm{eff} \simeq 0.4$ ($M=300$) and
$\kappa_\mathrm{eff} \simeq 0.23$ ($M=500$). This considerable suppression relative to $M=100$
grows further with $\phi_f^0$ as the larger coils are increasingly expelled. The widening deviation below 
the no-exclusion diagonal with increasing $M$ highlights the size-selective filtering function of the gel.

At low free polymer concentration these opposite tendencies -- swelling and pore opening for small coils 
versus deswelling and pore constriction for large ones -- establish a positive feedback loop that amplifies 
size selectivity: Open pores admit more small coils (raising their $\kappa_\mathrm{eff}$), while
constricted pores further exclude large coils (lowering their $\kappa_\mathrm{eff}$). At higher 
concentrations, the external osmotic pressure dominates for all $M$ and every pore contracts, but the 
selectivity persists, since for smaller coils the pores remain wider, with $\kappa_\mathrm{eff}$ 
correspondingly larger.

\begin{figure}[h]
\includegraphics[width=\columnwidth,angle=0]{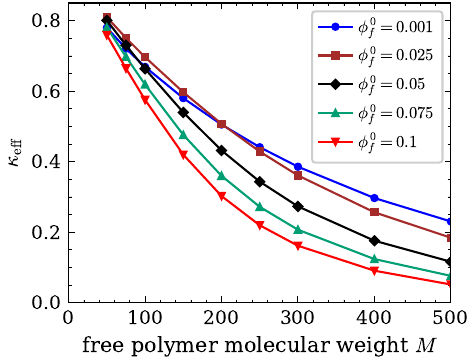}
\caption{Equilibrium partition coefficient, defined as free-polymer monomer volume fraction $\phi_f$ 
inside gel relative to reservoir value $\phi_f^0$ [equal to effective free-volume fraction 
$\kappa_\mathrm{eff}(L_\mathrm{eq}/R_g^0)$] versus coil molecular weight $M$ 
at reservoir concentrations $\phi_f^0=0.001$, $0.025$, $0.05$, $0.075$, and $0.1$.
The partition coefficient decreases monotonically with $M$, reflecting decreasing ratio of 
equilibrium pore size $L_\mathrm{eq}$ to reservoir coil size $R_g^0$. At fixed $M$, increasing $\phi_f^0$ 
reduces $\kappa_\mathrm{eff}$, as deswelling shrinks the pores. Curves are guides to the eye.}
\label{fig:K-vs-M}
\end{figure}

Figure~\ref{fig:K-vs-M} shows the equilibrium partition coefficient,
$\phi_f/\phi_f^0 = \kappa_\mathrm{eff}(L_\mathrm{eq}/R_g^0)$, 
as a function of free polymer molecular weight $M$ at several reservoir concentrations,
where $L_\mathrm{eq}$ is the equilibrium pore size. The partition coefficient decreases
monotonically with $M$, reflecting the increasing ratio of coil size to pore size. 
This behavior is analogous to the partition coefficient discussed by Zhou and Dill~\cite{ZhouDill2001} 
for a protein between two parallel walls [their Eq.~(13) and Fig.~6], which they showed decreases 
as the chain expands under denaturing conditions. In our model, $\kappa_{\rm eff}$ decreases as $M$ 
increases due to the same underlying physics (the ratio $R_g/L$ controls partitioning), but with 
molecular weight rather than solvent quality as the control variable.

In recent experiments, Takarai \textit{et al.}~\cite{Takarai2026} measured uptake of PEG by tetra-PEG gels
and demonstrated a universal partitioning law, $\phi_f/\phi_f^0 = \exp[-4(R_g/\ell_\mathrm{cycle})^2]$, 
where their mesh size $\ell_\mathrm{cycle}$ is proportional to our $L$. They found that the 
classical Ogston model underestimates the partitioning of flexible polymers into flexible gel networks. 
Our model predicts partition coefficients qualitatively consistent with, but quantitatively somewhat below,
the measurements of Takarai \textit{et al.}

\subsection{Crossover Molecular Weight}
\label{sec:results-crossover}
\begin{figure}[h]
\includegraphics[width=\columnwidth,angle=0]{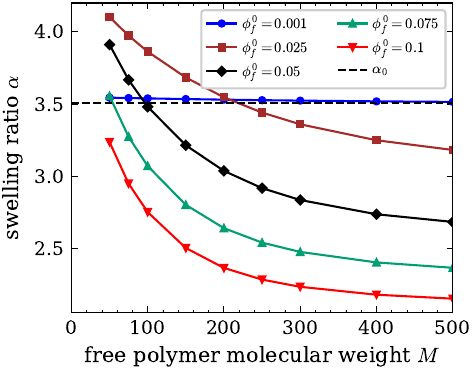}
\caption{Equilibrium swelling ratio $\alpha$ versus free polymer coil molecular weight $M$ 
at several reservoir concentrations $\phi_f^0$. With increasing $M$, larger coils induce deswelling
and $\alpha$ monotonically decreases. 
Crossover molecular weight at which $\alpha=\alpha_0$ (dashed line) decreases with 
increasing $\phi_f^0$. Curves are guides to the eye.}
\label{fig:alpha-vs-M}
\end{figure}

\begin{figure}[h]
\includegraphics[width=\columnwidth,angle=0]{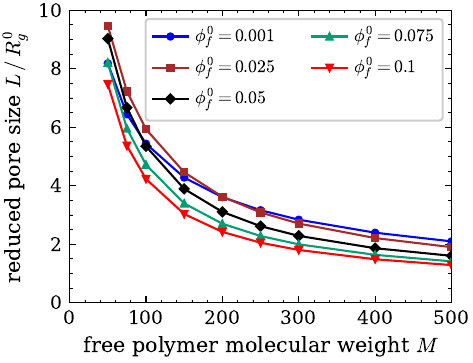}
\caption{Equilibrium reduced hydrogel pore size $L/R_g^0$ versus free polymer coil molecular weight $M$ 
at several reservoir concentrations $\phi_f^0$. Curves are guides to the eye.}
\label{fig:LRg-vs-M}
\end{figure}

Figure~\ref{fig:alpha-vs-M} shows our predictions for the equilibrium swelling ratio $\alpha$ 
as functions of free polymer coil molecular weight $M$ and reservoir concentration $\phi_f^0$ 
in the ranges $50\leq M\leq 500$ and $0.001\leq \phi_f^0\leq 0.1$. Most striking from these
data is that at all reservoir concentrations, $\alpha$ monotonically decreases with increasing $M$.
For $M<100$ and $\phi_f<0.05$, $\alpha$ exceeds its pure-solvent value $\alpha_0$, illustrating 
osmotic swelling induced by a low concentration of small, easily absorbed coils. In the most 
dilute case ($\phi_f^0=0.001$), $\alpha$ remains slightly above $\alpha_0$ over the whole $M$ range. 
For $\phi_f^0<0.05$, $\alpha$ starts above $\alpha_0$ at $M=50$ and monotonically decreases 
with increasing $M$, while at higher concentrations, $\alpha$ remains consistently below $\alpha_0$, 
reflecting deswelling induced by large, volume-excluded coils. Notably, the crossover molecular weight 
at which $\alpha=\alpha_0$, decreases with increasing $\phi_f^0$, as higher reservoir concentrations 
push the system into the deswelling regime at progressively smaller coil sizes.
At $\phi_f^0 = 0.1$, even $M = 100$ coils produce net deswelling.
Figure~\ref{fig:LRg-vs-M} shows the corresponding equilibrium reduced pore size $L/R_g^0$ as a 
function of molecular weight. With increasing $M$, as the hydrogel deswells, the pores contract,
increasingly so the higher the reservoir concentration.

Our results have direct implications for biopolymer folding. In particular, the confinement-induced 
reduction in equilibrium size and asphericity of absorbed free polymer is analogous to protein folding 
in which an extended, anisotropic (unfolded) conformation transitions to a compact, globular (folded) state.
Zhou and Dill~\cite{ZhouDill2001} predicted that confinement within a spherical cage can stabilize 
the folded state of a protein by more than $10\,k_BT$, an effect driven by the same selective elimination 
of extended conformations that are geometrically incompatible with the confining boundaries.
Our results provide a complementary, shape-resolved perspective on this mechanism. Confinement not only 
compresses the polymer (reducing $R_g$) but also reshapes it (reducing $A$), driving the conformational 
ensemble toward the globular geometry characteristic of natively folded proteins.

Crucially, the pore size is not a fixed parameter in the present model, but scales with the gel 
swelling ratio. Our results demonstrate that $\alpha$, and therefore $L$, responds sensitively to 
external conditions, including the concentration and molecular weight of free polymer in the reservoir.
This coupling establishes a direct, externally tunable link between the stimulus-responsive swelling 
of a hydrogel and the conformational state of any polymer confined within its pores. When the gel swells 
($\alpha$ increases and $L$ expands), the confinement relaxes and the embedded polymer adopts extended, 
high-$A$, unfolded conformations. On the other hand, when the gel deswells ($\alpha$ decreases and $L$ 
contracts), the confinement tightens and the polymer is driven toward compact, low-$A$, folded conformations.
This reversible, stimulus-controlled modulation of hydrogel swelling and polymer shape constitutes the 
central result of this work and provides a theoretical basis for using responsive microgels as synthetic, 
chaperone-like, nano-confinement platforms, whose folding action can be continuously tuned by external control.

\section{Conclusions}
\label{sec:conclusions}
In summary, we have developed a self-consistent mean-field theory for the equilibrium swelling and 
internal composition of a hydrogel mixed with a solution of free polymer coils.  
The theory extends the Flory-Rehner framework for polymer network swelling~\cite{FloryRehner1943a,FloryRehner1943b} 
and generalizes our previous work on microgel-nanoparticle mixtures~\cite{aryal2026response} by replacing rigid, 
hard-sphere crowders with flexible, self-avoiding-walk polymer coils modeled as fluctuating 
ellipsoids~\cite{Theodorou1985,rudnick1987shapes,lim-denton-JCP2016}.  
Conformational statistics of the free polymer coils are characterized by the eigenvalue distributions of the 
gyration tensor, and confinement within gel pores is captured through an effective free-volume fraction
$\kappa_\mathrm{eff}(L)$ parametrized from numerical free-volume theory modeling of SAW chains confined
between hard walls. Coupled equations expressing equality of chemical potentials inside and outside the
gel in chemical equilibrium are solved by numerical root-finding.

The central result of this work is that free polymer coils can cause a hydrogel to either swell or deswell, 
depending on their molecular weight and concentration. Relatively small coils readily penetrate gel pores 
and generate an internal osmotic pressure that drives solvent uptake and swelling, while larger coils
are substantially excluded from the gel interior, such that the external osmotic pressure of the reservoir 
dominates and compresses the gel. For intermediate free polymer molecular weights, the competition between these 
two effects produces a non-monotonic swelling response: initial swelling at low reservoir concentration 
followed by deswelling at higher concentration, in qualitative agreement with experimental observations 
of Vincent and coworkers~\cite{bradley-vincent2005,routh2006effect} for PNIPAM microgels in PEO solutions.

The gel effectively acts as a size-selective filter whose partitioning behavior is governed entirely by 
a single geometric factor, the effective free-volume fraction or partition coefficient. 
With increasing coil molecular weight, $\kappa_\mathrm{eff}$ decreases monotonically 
over the range studied, falling from 0.7 for $M=100$ to 0.25 for $M=500$ at $\phi_f^0=0.01$. 
This size-selective partitioning is consistent with the universal partitioning law recently established 
by Takarai \textit{et al.}~\cite{Takarai2026} for polymer chains in flexible gel networks.

Confinement within gel pores simultaneously decreases both the radius of gyration and the asphericity of 
embedded polymer coils, driving a transition from extended, prolate conformations toward compact, globular shapes, 
the polymer-physics analog of protein folding. This compaction is consistent with theoretical predictions 
that spatial confinement can stabilize folded protein states~\cite{ZhouDill2001,Zhou2008}.
Because the pore size scales with the swelling ratio ($L=L_0\alpha$), and the swelling ratio can respond to
external stimuli, such as temperature, pH, osmotic pressure, and free polymer concentration and molecular weight,
the conformational state of embedded polymers can be continuously and reversibly tuned. Gel swelling relaxes 
confinement and favors extended, unfolded states, while gel deswelling tightens confinement and promotes compact, 
folded states. This interplay establishes a quantitative, externally controllable connection between 
stimulus-responsive hydrogel swelling and biopolymer folding, providing a theoretical foundation for the 
design of artificial chaperone-like systems.

%resume
The present model makes several simplifying assumptions that suggest directions for future refinement. 
The Flory-Huggins interaction parameters were set to zero throughout, characteristic of good solvent
conditions. Incorporating non-ideal interactions ($\chi \neq 0$) would enable the study of 
temperature-dependent and solvent-quality effects on the coupled swelling-folding response, 
essential for modeling thermoresponsive PNIPAM-based systems near the volume phase transition. 

The parallel-wall confinement geometry is a simplified approximation of the complex, interconnected
pore network of a real hydrogel; more realistic pore geometries could be incorporated through modified
$\kappa_\mathrm{eff}$ parametrizations. The free polymer coils are treated as non-interacting. 
At higher internal concentrations, coil-coil excluded-volume interactions may become significant and 
could be incorporated through higher-order virial corrections. 

Finally, extending the present theory to concentrated suspensions of microgels via Monte Carlo simulations 
in the semi-grand canonical ensemble, incorporating microgel-microgel interactions via repulsive 
Hertzian pair potentials~\cite{Urich2016,zaccarelli2024}, would enable investigation of how crowding 
at the suspension level influences the single-particle swelling and confinement-induced folding predicted here.
The theory can also be extended to ionic hydrogels and microgels, in which electrostatic interactions can 
profoundly affect swelling~\cite{Weyer2018,Alziyadi2023,levin2002,colla-likos2014,moncho-jorda2013,
quesada-perez-moncho-jorda-anta2015,moncho-jorda-dzubiella2016}.

In conclusion, our results contribute to a growing understanding of how responsive soft materials 
can be engineered to control macromolecular conformations. By establishing a quantitative connection 
between externally tunable hydrogel swelling and the internal conformational state of embedded polymers, 
this work opens pathways toward applications in drug delivery, biosensing, tissue engineering, and the 
development of synthetic confinement platforms that replicate the chaperone function of biological cells.

\acknowledgments
\noindent This work was supported by the National Science Foundation under Grant No.~DMR-1928073.

\newpage
\begin{center}
{\bf AUTHOR DECLARATION}
\end{center}

\noindent {\bf Conflict of Interest} \\[1ex]
The authors have no conflicts to disclose.
\\[3ex]
{\bf Author Contributions} \\[2ex]
{\bf Mahesh Aryal:}
formal analysis (equal); methodology (equal); calculations (equal); software (equal);
data curation (equal); visualization (equal); validation (equal); writing -- original draft (equal);
writing -- review and editing (supporting).
\\[1ex]
{\bf Alan R.~Denton:}
conceptualization (lead); project administration (lead); resources (lead); supervision (lead);
formal analysis (equal); methodology (equal); calculations (equal); software (equal);
data curation (equal); visualization (equal); validation (equal); writing -- original draft (equal);
writing -- review and editing (lead).

\vspace*{0.3cm}
\begin{center}
{\bf DATA AVAILABILITY}
\end{center}

\noindent Data supporting the findings of this study are available from the corresponding author
upon reasonable request.

%%%%%%%%%%%%%%%%%%%%%%%%%%%%%%%%%%%%%%%%%%%%%%%%%%%%%%%%%%%%%%%%%%%%%%%%%%%
%%%%%%%                       REFERENCES
%%%%%%%%%%%%%%%%%%%%%%%%%%%%%%%%%%%%%%%%%%%%%%%%%%%%%%%%%%%%%%%%%%%%%%%%%%%

\bibliography{microgels}

\begin{thebibliography}{78}%
\makeatletter
\providecommand \@ifxundefined [1]{%
 \@ifx{#1\undefined}
}%
\providecommand \@ifnum [1]{%
 \ifnum #1\expandafter \@firstoftwo
 \else \expandafter \@secondoftwo
 \fi
}%
\providecommand \@ifx [1]{%
 \ifx #1\expandafter \@firstoftwo
 \else \expandafter \@secondoftwo
 \fi
}%
\providecommand \natexlab [1]{#1}%
\providecommand \enquote  [1]{``#1''}%
\providecommand \bibnamefont  [1]{#1}%
\providecommand \bibfnamefont [1]{#1}%
\providecommand \citenamefont [1]{#1}%
\providecommand \href@noop [0]{\@secondoftwo}%
\providecommand \href [0]{\begingroup \@sanitize@url \@href}%
\providecommand \@href[1]{\@@startlink{#1}\@@href}%
\providecommand \@@href[1]{\endgroup#1\@@endlink}%
\providecommand \@sanitize@url [0]{\catcode `\\12\catcode `\$12\catcode
  `\&12\catcode `\#12\catcode `\^12\catcode `\_12\catcode `\%12\relax}%
\providecommand \@@startlink[1]{}%
\providecommand \@@endlink[0]{}%
\providecommand \url  [0]{\begingroup\@sanitize@url \@url }%
\providecommand \@url [1]{\endgroup\@href {#1}{\urlprefix }}%
\providecommand \urlprefix  [0]{URL }%
\providecommand \Eprint [0]{\href }%
\providecommand \doibase [0]{http://dx.doi.org/}%
\providecommand \selectlanguage [0]{\@gobble}%
\providecommand \bibinfo  [0]{\@secondoftwo}%
\providecommand \bibfield  [0]{\@secondoftwo}%
\providecommand \translation [1]{[#1]}%
\providecommand \BibitemOpen [0]{}%
\providecommand \bibitemStop [0]{}%
\providecommand \bibitemNoStop [0]{.\EOS\space}%
\providecommand \EOS [0]{\spacefactor3000\relax}%
\providecommand \BibitemShut  [1]{\csname bibitem#1\endcsname}%
\let\auto@bib@innerbib\@empty
%</preamble>
\bibitem [{\citenamefont {Ellis}(2001)}]{Ellis2001}%
  \BibitemOpen
  \bibfield  {author} {\bibinfo {author} {\bibfnamefont {R.~J.}\ \bibnamefont
  {Ellis}},\ }\href {\doibase 10.1016/S0968-0004(01)01938-7} {\bibfield
  {journal} {\bibinfo  {journal} {Trends Biochem. Sci.}\ }\textbf {\bibinfo
  {volume} {26}},\ \bibinfo {pages} {597} (\bibinfo {year} {2001})}\BibitemShut
  {NoStop}%
\bibitem [{\citenamefont {Zimmerman}\ and\ \citenamefont
  {Minton}(1993)}]{Zimmerman1991}%
  \BibitemOpen
  \bibfield  {author} {\bibinfo {author} {\bibfnamefont {S.~B.}\ \bibnamefont
  {Zimmerman}}\ and\ \bibinfo {author} {\bibfnamefont {A.~P.}\ \bibnamefont
  {Minton}},\ }\href {\doibase 10.1146/annurev.bb.22.060193.000331} {\bibfield
  {journal} {\bibinfo  {journal} {Annu. Rev. Biophys. Biomol. Struct.}\
  }\textbf {\bibinfo {volume} {22}},\ \bibinfo {pages} {27} (\bibinfo {year}
  {1993})}\BibitemShut {NoStop}%
\bibitem [{\citenamefont {Hancock}(2012)}]{hancock2012}%
  \BibitemOpen
  \bibfield  {author} {\bibinfo {author} {\bibfnamefont {R.}~\bibnamefont
  {Hancock}},\ }in\ \href@noop {} {\emph {\bibinfo {booktitle} {Genome
  Organization and Function in the Cell Nucleus}}},\ \bibinfo {editor} {edited
  by\ \bibinfo {editor} {\bibfnamefont {K.}~\bibnamefont {Rippe}}}\ (\bibinfo
  {publisher} {Wiley-VCH},\ \bibinfo {address} {Weinheim},\ \bibinfo {year}
  {2012})\ pp.\ \bibinfo {pages} {169--184}\BibitemShut {NoStop}%
\bibitem [{\citenamefont {Denton}(2013)}]{denton-cmb2013}%
  \BibitemOpen
  \bibfield  {author} {\bibinfo {author} {\bibfnamefont {A.~R.}\ \bibnamefont
  {Denton}},\ }in\ \href@noop {} {\emph {\bibinfo {booktitle} {New Models of
  the Cell Nucleus: Crowding and Entropic Forces and Phase Separation and
  Fractals}}},\ \bibinfo {editor} {edited by\ \bibinfo {editor} {\bibfnamefont
  {R.}~\bibnamefont {Hancock}}\ and\ \bibinfo {editor} {\bibfnamefont {K.~W.}\
  \bibnamefont {Jeon}}}\ (\bibinfo  {publisher} {Academic Press},\ \bibinfo
  {address} {UK},\ \bibinfo {year} {2013})\ pp.\ \bibinfo {pages}
  {27--72}\BibitemShut {NoStop}%
\bibitem [{\citenamefont {Minton}(2008)}]{Minton2008}%
  \BibitemOpen
  \bibfield  {author} {\bibinfo {author} {\bibfnamefont {A.~P.}\ \bibnamefont
  {Minton}},\ }\href@noop {} {\bibfield  {journal} {\bibinfo  {journal}
  {Biophys. Chem.}\ }\textbf {\bibinfo {volume} {119}},\ \bibinfo {pages} {271}
  (\bibinfo {year} {2008})}\BibitemShut {NoStop}%
\bibitem [{\citenamefont {Zhou}\ \emph {et~al.}(2008)\citenamefont {Zhou},
  \citenamefont {Rivas},\ and\ \citenamefont {Minton}}]{Zhou2008review}%
  \BibitemOpen
  \bibfield  {author} {\bibinfo {author} {\bibfnamefont {H.-X.}\ \bibnamefont
  {Zhou}}, \bibinfo {author} {\bibfnamefont {G.}~\bibnamefont {Rivas}}, \ and\
  \bibinfo {author} {\bibfnamefont {A.~P.}\ \bibnamefont {Minton}},\ }\href
  {\doibase 10.1146/annurev.biophys.37.032807.125817} {\bibfield  {journal}
  {\bibinfo  {journal} {Annu. Rev. Biophys.}\ }\textbf {\bibinfo {volume}
  {37}},\ \bibinfo {pages} {375} (\bibinfo {year} {2008})}\BibitemShut
  {NoStop}%
\bibitem [{\citenamefont {Cheung}\ \emph {et~al.}(2005)\citenamefont {Cheung},
  \citenamefont {Klimov},\ and\ \citenamefont {Thirumalai}}]{Cheung2005}%
  \BibitemOpen
  \bibfield  {author} {\bibinfo {author} {\bibfnamefont {M.~S.}\ \bibnamefont
  {Cheung}}, \bibinfo {author} {\bibfnamefont {D.}~\bibnamefont {Klimov}}, \
  and\ \bibinfo {author} {\bibfnamefont {D.}~\bibnamefont {Thirumalai}},\
  }\href {\doibase 10.1073/pnas.0409630102} {\bibfield  {journal} {\bibinfo
  {journal} {Proc. Natl. Acad. Sci.}\ }\textbf {\bibinfo {volume} {102}},\
  \bibinfo {pages} {4753} (\bibinfo {year} {2005})}\BibitemShut {NoStop}%
\bibitem [{\citenamefont {Koldewey}\ \emph {et~al.}(2017)\citenamefont
  {Koldewey}, \citenamefont {Horowitz},\ and\ \citenamefont
  {Bardwell}}]{Koldewey2017}%
  \BibitemOpen
  \bibfield  {author} {\bibinfo {author} {\bibfnamefont {P.}~\bibnamefont
  {Koldewey}}, \bibinfo {author} {\bibfnamefont {S.}~\bibnamefont {Horowitz}},
  \ and\ \bibinfo {author} {\bibfnamefont {J.~C.}\ \bibnamefont {Bardwell}},\
  }\href {\doibase 10.1074/jbc.R117.789107} {\bibfield  {journal} {\bibinfo
  {journal} {J. Biol. Chem.}\ }\textbf {\bibinfo {volume} {292}},\ \bibinfo
  {pages} {12479} (\bibinfo {year} {2017})}\BibitemShut {NoStop}%
\bibitem [{\citenamefont {Hartl}\ and\ \citenamefont
  {Hayer-Hartl}(2002)}]{Hartl2002}%
  \BibitemOpen
  \bibfield  {author} {\bibinfo {author} {\bibfnamefont {F.~U.}\ \bibnamefont
  {Hartl}}\ and\ \bibinfo {author} {\bibfnamefont {M.}~\bibnamefont
  {Hayer-Hartl}},\ }\href {\doibase 10.1126/science.1068408} {\bibfield
  {journal} {\bibinfo  {journal} {Science}\ }\textbf {\bibinfo {volume}
  {295}},\ \bibinfo {pages} {1852} (\bibinfo {year} {2002})}\BibitemShut
  {NoStop}%
\bibitem [{\citenamefont {Ziv}\ \emph {et~al.}(2005)\citenamefont {Ziv},
  \citenamefont {Haran},\ and\ \citenamefont {Thirumalai}}]{Ziv2005}%
  \BibitemOpen
  \bibfield  {author} {\bibinfo {author} {\bibfnamefont {G.}~\bibnamefont
  {Ziv}}, \bibinfo {author} {\bibfnamefont {G.}~\bibnamefont {Haran}}, \ and\
  \bibinfo {author} {\bibfnamefont {D.}~\bibnamefont {Thirumalai}},\ }\href
  {\doibase 10.1073/pnas.0508234102} {\bibfield  {journal} {\bibinfo  {journal}
  {Proc. Natl. Acad. Sci.}\ }\textbf {\bibinfo {volume} {102}},\ \bibinfo
  {pages} {18956} (\bibinfo {year} {2005})}\BibitemShut {NoStop}%
\bibitem [{\citenamefont {Zhou}\ and\ \citenamefont
  {Dill}(2001)}]{ZhouDill2001}%
  \BibitemOpen
  \bibfield  {author} {\bibinfo {author} {\bibfnamefont {H.-X.}\ \bibnamefont
  {Zhou}}\ and\ \bibinfo {author} {\bibfnamefont {K.~A.}\ \bibnamefont
  {Dill}},\ }\href {\doibase 10.1021/bi0155504} {\bibfield  {journal} {\bibinfo
   {journal} {Biochem.}\ }\textbf {\bibinfo {volume} {40}},\ \bibinfo {pages}
  {11289} (\bibinfo {year} {2001})}\BibitemShut {NoStop}%
\bibitem [{\citenamefont {Ping}\ \emph {et~al.}(2004)\citenamefont {Ping},
  \citenamefont {Yuan}, \citenamefont {Sun},\ and\ \citenamefont
  {Wei}}]{Ping2004}%
  \BibitemOpen
  \bibfield  {author} {\bibinfo {author} {\bibfnamefont {G.}~\bibnamefont
  {Ping}}, \bibinfo {author} {\bibfnamefont {J.-M.}\ \bibnamefont {Yuan}},
  \bibinfo {author} {\bibfnamefont {Z.}~\bibnamefont {Sun}}, \ and\ \bibinfo
  {author} {\bibfnamefont {Y.}~\bibnamefont {Wei}},\ }\href {\doibase
  10.1002/jmr.710} {\bibfield  {journal} {\bibinfo  {journal} {J. Mol.
  Recognit.}\ }\textbf {\bibinfo {volume} {17}},\ \bibinfo {pages} {433}
  (\bibinfo {year} {2004})}\BibitemShut {NoStop}%
\bibitem [{\citenamefont {Klimov}\ \emph {et~al.}(2002)\citenamefont {Klimov},
  \citenamefont {Newfield},\ and\ \citenamefont {Thirumalai}}]{Klimov2002}%
  \BibitemOpen
  \bibfield  {author} {\bibinfo {author} {\bibfnamefont {D.~K.}\ \bibnamefont
  {Klimov}}, \bibinfo {author} {\bibfnamefont {D.}~\bibnamefont {Newfield}}, \
  and\ \bibinfo {author} {\bibfnamefont {D.}~\bibnamefont {Thirumalai}},\
  }\href {\doibase 10.1073/pnas.072220699} {\bibfield  {journal} {\bibinfo
  {journal} {Proc. Natl. Acad. Sci.}\ }\textbf {\bibinfo {volume} {99}},\
  \bibinfo {pages} {8019} (\bibinfo {year} {2002})}\BibitemShut {NoStop}%
\bibitem [{\citenamefont {Zhou}(2008)}]{Zhou2008}%
  \BibitemOpen
  \bibfield  {author} {\bibinfo {author} {\bibfnamefont {H.-X.}\ \bibnamefont
  {Zhou}},\ }\href {\doibase 10.1016/j.abb.2007.07.013} {\bibfield  {journal}
  {\bibinfo  {journal} {Arch. Biochem. Biophys.}\ }\textbf {\bibinfo {volume}
  {469}},\ \bibinfo {pages} {76} (\bibinfo {year} {2008})}\BibitemShut
  {NoStop}%
\bibitem [{\citenamefont {Eggers}\ and\ \citenamefont
  {Valentine}(2001)}]{Eggers2001}%
  \BibitemOpen
  \bibfield  {author} {\bibinfo {author} {\bibfnamefont {D.~K.}\ \bibnamefont
  {Eggers}}\ and\ \bibinfo {author} {\bibfnamefont {J.~S.}\ \bibnamefont
  {Valentine}},\ }\href {\doibase 10.1110/ps.36201} {\bibfield  {journal}
  {\bibinfo  {journal} {Protein Science}\ }\textbf {\bibinfo {volume} {10}},\
  \bibinfo {pages} {250} (\bibinfo {year} {2001})}\BibitemShut {NoStop}%
\bibitem [{\citenamefont {Zhou}(2004)}]{Zhou2004}%
  \BibitemOpen
  \bibfield  {author} {\bibinfo {author} {\bibfnamefont {H.-X.}\ \bibnamefont
  {Zhou}},\ }\href {\doibase 10.1002/jmr.711} {\bibfield  {journal} {\bibinfo
  {journal} {J. Mol. Recognit.}\ }\textbf {\bibinfo {volume} {17}},\ \bibinfo
  {pages} {535} (\bibinfo {year} {2004})}\BibitemShut {NoStop}%
\bibitem [{\citenamefont {Minton}(2005)}]{Minton2005}%
  \BibitemOpen
  \bibfield  {author} {\bibinfo {author} {\bibfnamefont {A.~P.}\ \bibnamefont
  {Minton}},\ }\href {\doibase 10.1529/biophysj.104.050351} {\bibfield
  {journal} {\bibinfo  {journal} {Biophys. J.}\ }\textbf {\bibinfo {volume}
  {88}},\ \bibinfo {pages} {971} (\bibinfo {year} {2005})}\BibitemShut
  {NoStop}%
\bibitem [{\citenamefont {Ross}\ \emph {et~al.}(2020)\citenamefont {Ross},
  \citenamefont {Kunkel}, \citenamefont {Long},\ and\ \citenamefont
  {Asuri}}]{ross2020combined}%
  \BibitemOpen
  \bibfield  {author} {\bibinfo {author} {\bibfnamefont {M.~L.}\ \bibnamefont
  {Ross}}, \bibinfo {author} {\bibfnamefont {J.}~\bibnamefont {Kunkel}},
  \bibinfo {author} {\bibfnamefont {S.}~\bibnamefont {Long}}, \ and\ \bibinfo
  {author} {\bibfnamefont {P.}~\bibnamefont {Asuri}},\ }\href@noop {}
  {\bibfield  {journal} {\bibinfo  {journal} {Int. J. Mol. Sci.}\ }\textbf
  {\bibinfo {volume} {21}},\ \bibinfo {pages} {8516} (\bibinfo {year}
  {2020})}\BibitemShut {NoStop}%
\bibitem [{\citenamefont {Lyon}\ and\ \citenamefont
  {Serpe}(2012)}]{HydrogelBook2012}%
  \BibitemOpen
  \bibinfo {editor} {\bibfnamefont {L.~A.}\ \bibnamefont {Lyon}}\ and\ \bibinfo
  {editor} {\bibfnamefont {M.~J.}\ \bibnamefont {Serpe}},\ eds.,\ \href@noop {}
  {\emph {\bibinfo {title} {Hydrogel Micro and Nanoparticles}}}\ (\bibinfo
  {publisher} {Wiley-VCH Verlag GmbH \& Co. KGaA},\ \bibinfo {address}
  {Weinheim},\ \bibinfo {year} {2012})\BibitemShut {NoStop}%
\bibitem [{\citenamefont {Peppas}\ \emph {et~al.}(2006)\citenamefont {Peppas},
  \citenamefont {Hilt}, \citenamefont {Khademhosseini},\ and\ \citenamefont
  {Langer}}]{Peppas2006}%
  \BibitemOpen
  \bibfield  {author} {\bibinfo {author} {\bibfnamefont {N.~A.}\ \bibnamefont
  {Peppas}}, \bibinfo {author} {\bibfnamefont {J.~Z.}\ \bibnamefont {Hilt}},
  \bibinfo {author} {\bibfnamefont {A.}~\bibnamefont {Khademhosseini}}, \ and\
  \bibinfo {author} {\bibfnamefont {R.}~\bibnamefont {Langer}},\ }\href
  {\doibase 10.1002/adma.200501612} {\bibfield  {journal} {\bibinfo  {journal}
  {Adv. Mat.}\ }\textbf {\bibinfo {volume} {18}},\ \bibinfo {pages} {1345}
  (\bibinfo {year} {2006})}\BibitemShut {NoStop}%
\bibitem [{\citenamefont {Kaberova}\ \emph {et~al.}(2020)\citenamefont
  {Kaberova}, \citenamefont {Karpushkin}, \citenamefont {Nevoralová},
  \citenamefont {Vetrík}, \citenamefont {Šlouf},\ and\ \citenamefont
  {Dušková-Smrčková}}]{kaberova2020microscopic}%
  \BibitemOpen
  \bibfield  {author} {\bibinfo {author} {\bibfnamefont {Z.}~\bibnamefont
  {Kaberova}}, \bibinfo {author} {\bibfnamefont {E.}~\bibnamefont
  {Karpushkin}}, \bibinfo {author} {\bibfnamefont {M.}~\bibnamefont
  {Nevoralová}}, \bibinfo {author} {\bibfnamefont {M.}~\bibnamefont
  {Vetrík}}, \bibinfo {author} {\bibfnamefont {M.}~\bibnamefont {Šlouf}}, \
  and\ \bibinfo {author} {\bibfnamefont {M.}~\bibnamefont
  {Dušková-Smrčková}},\ }\href@noop {} {\bibfield  {journal} {\bibinfo
  {journal} {Polymers}\ }\textbf {\bibinfo {volume} {12}},\ \bibinfo {pages}
  {578} (\bibinfo {year} {2020})}\BibitemShut {NoStop}%
\bibitem [{\citenamefont {Pelton}(2000)}]{Pelton2000}%
  \BibitemOpen
  \bibfield  {author} {\bibinfo {author} {\bibfnamefont {R.}~\bibnamefont
  {Pelton}},\ }\href {\doibase 10.1016/S0001-8686(99)00023-8} {\bibfield
  {journal} {\bibinfo  {journal} {Adv. Colloid Interface Sci.}\ }\textbf
  {\bibinfo {volume} {85}},\ \bibinfo {pages} {1} (\bibinfo {year}
  {2000})}\BibitemShut {NoStop}%
\bibitem [{\citenamefont {Fern{\'a}ndez-Nieves}\ \emph
  {et~al.}(2011)\citenamefont {Fern{\'a}ndez-Nieves}, \citenamefont {Wyss},
  \citenamefont {Mattsson},\ and\ \citenamefont {Weitz}}]{MicrogelBook2011}%
  \BibitemOpen
  \bibinfo {editor} {\bibfnamefont {A.}~\bibnamefont {Fern{\'a}ndez-Nieves}},
  \bibinfo {editor} {\bibfnamefont {H.}~\bibnamefont {Wyss}}, \bibinfo {editor}
  {\bibfnamefont {J.}~\bibnamefont {Mattsson}}, \ and\ \bibinfo {editor}
  {\bibfnamefont {D.~A.}\ \bibnamefont {Weitz}},\ eds.,\ \href@noop {} {\emph
  {\bibinfo {title} {Microgel Suspensions: Fundamentals and Applications}}}\
  (\bibinfo  {publisher} {Wiley-VCH Verlag GmbH \& Co. KGaA},\ \bibinfo
  {address} {Weinheim},\ \bibinfo {year} {2011})\BibitemShut {NoStop}%
\bibitem [{\citenamefont {Riest}\ \emph {et~al.}(2012)\citenamefont {Riest},
  \citenamefont {Mohanty}, \citenamefont {Schurtenberger},\ and\ \citenamefont
  {Likos}}]{schurtenberger-ZPC2012}%
  \BibitemOpen
  \bibfield  {author} {\bibinfo {author} {\bibfnamefont {J.}~\bibnamefont
  {Riest}}, \bibinfo {author} {\bibfnamefont {P.}~\bibnamefont {Mohanty}},
  \bibinfo {author} {\bibfnamefont {P.}~\bibnamefont {Schurtenberger}}, \ and\
  \bibinfo {author} {\bibfnamefont {C.~N.}\ \bibnamefont {Likos}},\ }\href@noop
  {} {\bibfield  {journal} {\bibinfo  {journal} {Z. Phys. Chem.}\ }\textbf
  {\bibinfo {volume} {226}},\ \bibinfo {pages} {711} (\bibinfo {year}
  {2012})}\BibitemShut {NoStop}%
\bibitem [{\citenamefont {Colla}\ and\ \citenamefont
  {Likos}(2015)}]{Colla2015}%
  \BibitemOpen
  \bibfield  {author} {\bibinfo {author} {\bibfnamefont {T.}~\bibnamefont
  {Colla}}\ and\ \bibinfo {author} {\bibfnamefont {C.~N.}\ \bibnamefont
  {Likos}},\ }\href {\doibase 10.1080/00268976.2015.1005194} {\bibfield
  {journal} {\bibinfo  {journal} {Mol. Phys.}\ }\textbf {\bibinfo {volume}
  {113}},\ \bibinfo {pages} {2496} (\bibinfo {year} {2015})}\BibitemShut
  {NoStop}%
\bibitem [{\citenamefont {Plamper}\ and\ \citenamefont
  {Richtering}(2017)}]{plamper2017functional}%
  \BibitemOpen
  \bibfield  {author} {\bibinfo {author} {\bibfnamefont {F.~A.}\ \bibnamefont
  {Plamper}}\ and\ \bibinfo {author} {\bibfnamefont {W.}~\bibnamefont
  {Richtering}},\ }\href@noop {} {\bibfield  {journal} {\bibinfo  {journal}
  {Acc. Chem. Res.}\ }\textbf {\bibinfo {volume} {50}},\ \bibinfo {pages} {131}
  (\bibinfo {year} {2017})}\BibitemShut {NoStop}%
\bibitem [{\citenamefont {Karg}\ \emph {et~al.}(2019)\citenamefont {Karg},
  \citenamefont {Pich}, \citenamefont {Hellweg}, \citenamefont {Hoare},
  \citenamefont {Lyon}, \citenamefont {Crassous}, \citenamefont {Suzuki},
  \citenamefont {Gumerov}, \citenamefont {Schneider}, \citenamefont
  {Potemkin},\ and\ \citenamefont {Richtering}}]{Karg2019}%
  \BibitemOpen
  \bibfield  {author} {\bibinfo {author} {\bibfnamefont {M.}~\bibnamefont
  {Karg}}, \bibinfo {author} {\bibfnamefont {A.}~\bibnamefont {Pich}}, \bibinfo
  {author} {\bibfnamefont {T.}~\bibnamefont {Hellweg}}, \bibinfo {author}
  {\bibfnamefont {T.}~\bibnamefont {Hoare}}, \bibinfo {author} {\bibfnamefont
  {L.~A.}\ \bibnamefont {Lyon}}, \bibinfo {author} {\bibfnamefont {J.~J.}\
  \bibnamefont {Crassous}}, \bibinfo {author} {\bibfnamefont {D.}~\bibnamefont
  {Suzuki}}, \bibinfo {author} {\bibfnamefont {R.~A.}\ \bibnamefont {Gumerov}},
  \bibinfo {author} {\bibfnamefont {S.}~\bibnamefont {Schneider}}, \bibinfo
  {author} {\bibfnamefont {I.~I.}\ \bibnamefont {Potemkin}}, \ and\ \bibinfo
  {author} {\bibfnamefont {W.}~\bibnamefont {Richtering}},\ }\href {\doibase
  10.1021/acs.langmuir.8b04304} {\bibfield  {journal} {\bibinfo  {journal}
  {Langmuir}\ }\textbf {\bibinfo {volume} {35}},\ \bibinfo {pages} {6231}
  (\bibinfo {year} {2019})}\BibitemShut {NoStop}%
\bibitem [{\citenamefont {Scheffold}(2020)}]{scheffold2020pathways}%
  \BibitemOpen
  \bibfield  {author} {\bibinfo {author} {\bibfnamefont {F.}~\bibnamefont
  {Scheffold}},\ }\href@noop {} {\bibfield  {journal} {\bibinfo  {journal}
  {Nat. Commun.}\ }\textbf {\bibinfo {volume} {11}},\ \bibinfo {pages} {4315}
  (\bibinfo {year} {2020})}\BibitemShut {NoStop}%
\bibitem [{\citenamefont {Scotti}\ \emph {et~al.}(2022)\citenamefont {Scotti},
  \citenamefont {Schulte}, \citenamefont {Lopez}, \citenamefont {Crassous},
  \citenamefont {Bochenek},\ and\ \citenamefont
  {Richtering}}]{Scotti2022review}%
  \BibitemOpen
  \bibfield  {author} {\bibinfo {author} {\bibfnamefont {A.}~\bibnamefont
  {Scotti}}, \bibinfo {author} {\bibfnamefont {M.~F.}\ \bibnamefont {Schulte}},
  \bibinfo {author} {\bibfnamefont {C.~G.}\ \bibnamefont {Lopez}}, \bibinfo
  {author} {\bibfnamefont {J.~J.}\ \bibnamefont {Crassous}}, \bibinfo {author}
  {\bibfnamefont {S.}~\bibnamefont {Bochenek}}, \ and\ \bibinfo {author}
  {\bibfnamefont {W.}~\bibnamefont {Richtering}},\ }\href {\doibase
  10.1021/acs.chemrev.2c00035} {\bibfield  {journal} {\bibinfo  {journal}
  {Chem. Rev.}\ }\textbf {\bibinfo {volume} {122}},\ \bibinfo {pages} {11675}
  (\bibinfo {year} {2022})}\BibitemShut {NoStop}%
\bibitem [{\citenamefont {Del~Monte}\ and\ \citenamefont
  {Zaccarelli}(2024)}]{zaccarelli2024}%
  \BibitemOpen
  \bibfield  {author} {\bibinfo {author} {\bibfnamefont {G.}~\bibnamefont
  {Del~Monte}}\ and\ \bibinfo {author} {\bibfnamefont {E.}~\bibnamefont
  {Zaccarelli}},\ }\href@noop {} {\bibfield  {journal} {\bibinfo  {journal}
  {Phys. Rev. X}\ }\textbf {\bibinfo {volume} {14}},\ \bibinfo {pages} {041067}
  (\bibinfo {year} {2024})}\BibitemShut {NoStop}%
\bibitem [{\citenamefont {Moncho-Jord{\'a}}\ \emph {et~al.}(2025)\citenamefont
  {Moncho-Jord{\'a}}, \citenamefont {Cuetos}, \citenamefont
  {Fernandez-Rodriguez}, \citenamefont {Dzubiella},\ and\ \citenamefont
  {Patti}}]{moncho-jorda2025}%
  \BibitemOpen
  \bibfield  {author} {\bibinfo {author} {\bibfnamefont {A.}~\bibnamefont
  {Moncho-Jord{\'a}}}, \bibinfo {author} {\bibfnamefont {A.}~\bibnamefont
  {Cuetos}}, \bibinfo {author} {\bibfnamefont {M.~A.}\ \bibnamefont
  {Fernandez-Rodriguez}}, \bibinfo {author} {\bibfnamefont {J.}~\bibnamefont
  {Dzubiella}}, \ and\ \bibinfo {author} {\bibfnamefont {A.}~\bibnamefont
  {Patti}},\ }\href@noop {} {\bibfield  {journal} {\bibinfo  {journal}
  {Macromolecules}\ }\textbf {\bibinfo {volume} {58}},\ \bibinfo {pages}
  {10659} (\bibinfo {year} {2025})}\BibitemShut {NoStop}%
\bibitem [{\citenamefont {Flory}\ and\ \citenamefont
  {Rehner}(1943{\natexlab{a}})}]{FloryRehner1943a}%
  \BibitemOpen
  \bibfield  {author} {\bibinfo {author} {\bibfnamefont {P.~J.}\ \bibnamefont
  {Flory}}\ and\ \bibinfo {author} {\bibfnamefont {J.}~\bibnamefont {Rehner}},\
  }\href {\doibase 10.1063/1.1723791} {\bibfield  {journal} {\bibinfo
  {journal} {J. Chem. Phys.}\ }\textbf {\bibinfo {volume} {11}},\ \bibinfo
  {pages} {512} (\bibinfo {year} {1943}{\natexlab{a}})}\BibitemShut {NoStop}%
\bibitem [{\citenamefont {Flory}\ and\ \citenamefont
  {Rehner}(1943{\natexlab{b}})}]{FloryRehner1943b}%
  \BibitemOpen
  \bibfield  {author} {\bibinfo {author} {\bibfnamefont {P.~J.}\ \bibnamefont
  {Flory}}\ and\ \bibinfo {author} {\bibfnamefont {J.}~\bibnamefont {Rehner}},\
  }\href {\doibase 10.1063/1.1723792} {\bibfield  {journal} {\bibinfo
  {journal} {J. Chem. Phys.}\ }\textbf {\bibinfo {volume} {11}},\ \bibinfo
  {pages} {521} (\bibinfo {year} {1943}{\natexlab{b}})}\BibitemShut {NoStop}%
\bibitem [{\citenamefont {Flory}(1953)}]{Flory}%
  \BibitemOpen
  \bibfield  {author} {\bibinfo {author} {\bibfnamefont {P.~J.}\ \bibnamefont
  {Flory}},\ }\href@noop {} {\emph {\bibinfo {title} {Principles of Polymer
  Chemistry}}}\ (\bibinfo  {publisher} {Cornell University Press},\ \bibinfo
  {address} {Ithaca},\ \bibinfo {year} {1953})\BibitemShut {NoStop}%
\bibitem [{\citenamefont {Erman}\ and\ \citenamefont
  {Flory}(1986)}]{Erman1986}%
  \BibitemOpen
  \bibfield  {author} {\bibinfo {author} {\bibfnamefont {B.}~\bibnamefont
  {Erman}}\ and\ \bibinfo {author} {\bibfnamefont {P.~J.}\ \bibnamefont
  {Flory}},\ }\href {\doibase 10.1021/ma00163a003} {\bibfield  {journal}
  {\bibinfo  {journal} {Macromol.}\ }\textbf {\bibinfo {volume} {19}},\
  \bibinfo {pages} {2342} (\bibinfo {year} {1986})}\BibitemShut {NoStop}%
\bibitem [{\citenamefont {Lopez}\ and\ \citenamefont
  {Richtering}(2017)}]{Lopez2017}%
  \BibitemOpen
  \bibfield  {author} {\bibinfo {author} {\bibfnamefont {C.~G.}\ \bibnamefont
  {Lopez}}\ and\ \bibinfo {author} {\bibfnamefont {W.}~\bibnamefont
  {Richtering}},\ }\href {\doibase 10.1039/C7SM01274H} {\bibfield  {journal}
  {\bibinfo  {journal} {Soft Matter}\ }\textbf {\bibinfo {volume} {13}},\
  \bibinfo {pages} {8271} (\bibinfo {year} {2017})}\BibitemShut {NoStop}%
\bibitem [{\citenamefont {Friesen}\ \emph {et~al.}(2022)\citenamefont
  {Friesen}, \citenamefont {Hannappel}, \citenamefont {Kakorin},\ and\
  \citenamefont {Hellweg}}]{Friesen2022}%
  \BibitemOpen
  \bibfield  {author} {\bibinfo {author} {\bibfnamefont {S.}~\bibnamefont
  {Friesen}}, \bibinfo {author} {\bibfnamefont {Y.}~\bibnamefont {Hannappel}},
  \bibinfo {author} {\bibfnamefont {S.}~\bibnamefont {Kakorin}}, \ and\
  \bibinfo {author} {\bibfnamefont {T.}~\bibnamefont {Hellweg}},\ }\href
  {\doibase 10.1007/s00396-022-04950-w} {\bibfield  {journal} {\bibinfo
  {journal} {Colloid Polym. Sci.}\ }\textbf {\bibinfo {volume} {300}},\
  \bibinfo {pages} {1123} (\bibinfo {year} {2022})}\BibitemShut {NoStop}%
\bibitem [{\citenamefont {Urich}\ and\ \citenamefont
  {Denton}(2016)}]{Urich2016}%
  \BibitemOpen
  \bibfield  {author} {\bibinfo {author} {\bibfnamefont {M.}~\bibnamefont
  {Urich}}\ and\ \bibinfo {author} {\bibfnamefont {A.~R.}\ \bibnamefont
  {Denton}},\ }\href {\doibase 10.1039/C6SM02056A} {\bibfield  {journal}
  {\bibinfo  {journal} {Soft Matter}\ }\textbf {\bibinfo {volume} {12}},\
  \bibinfo {pages} {9086} (\bibinfo {year} {2016})}\BibitemShut {NoStop}%
\bibitem [{\citenamefont {Weyer}\ and\ \citenamefont
  {Denton}(2018)}]{Weyer2018}%
  \BibitemOpen
  \bibfield  {author} {\bibinfo {author} {\bibfnamefont {T.~J.}\ \bibnamefont
  {Weyer}}\ and\ \bibinfo {author} {\bibfnamefont {A.~R.}\ \bibnamefont
  {Denton}},\ }\href {\doibase 10.1039/C8SM00799C} {\bibfield  {journal}
  {\bibinfo  {journal} {Soft Matter}\ }\textbf {\bibinfo {volume} {14}},\
  \bibinfo {pages} {4530} (\bibinfo {year} {2018})}\BibitemShut {NoStop}%
\bibitem [{\citenamefont {Alziyadi}\ and\ \citenamefont
  {Denton}(2023)}]{Alziyadi2023}%
  \BibitemOpen
  \bibfield  {author} {\bibinfo {author} {\bibfnamefont {M.~O.}\ \bibnamefont
  {Alziyadi}}\ and\ \bibinfo {author} {\bibfnamefont {A.~R.}\ \bibnamefont
  {Denton}},\ }\href@noop {} {\bibfield  {journal} {\bibinfo  {journal} {J.
  Chem. Phys.}\ }\textbf {\bibinfo {volume} {159}} (\bibinfo {year}
  {2023})}\BibitemShut {NoStop}%
\bibitem [{\citenamefont {Brito}\ \emph {et~al.}(2019)\citenamefont {Brito},
  \citenamefont {Denton},\ and\ \citenamefont {Nägele}}]{Brito2019}%
  \BibitemOpen
  \bibfield  {author} {\bibinfo {author} {\bibfnamefont {M.~E.}\ \bibnamefont
  {Brito}}, \bibinfo {author} {\bibfnamefont {A.~R.}\ \bibnamefont {Denton}}, \
  and\ \bibinfo {author} {\bibfnamefont {G.}~\bibnamefont {Nägele}},\ }\href
  {\doibase 10.1063/1.5129575} {\bibfield  {journal} {\bibinfo  {journal} {J.
  Chem. Phys.}\ }\textbf {\bibinfo {volume} {151}},\ \bibinfo {pages} {224901}
  (\bibinfo {year} {2019})}\BibitemShut {NoStop}%
\bibitem [{\citenamefont {Scotti}\ \emph {et~al.}(2019)\citenamefont {Scotti},
  \citenamefont {Denton}, \citenamefont {Brugnoni}, \citenamefont {Houston},
  \citenamefont {Schweins}, \citenamefont {Potemkin},\ and\ \citenamefont
  {Richtering}}]{Scotti2019}%
  \BibitemOpen
  \bibfield  {author} {\bibinfo {author} {\bibfnamefont {A.}~\bibnamefont
  {Scotti}}, \bibinfo {author} {\bibfnamefont {A.~R.}\ \bibnamefont {Denton}},
  \bibinfo {author} {\bibfnamefont {M.}~\bibnamefont {Brugnoni}}, \bibinfo
  {author} {\bibfnamefont {J.~E.}\ \bibnamefont {Houston}}, \bibinfo {author}
  {\bibfnamefont {R.}~\bibnamefont {Schweins}}, \bibinfo {author}
  {\bibfnamefont {I.~I.}\ \bibnamefont {Potemkin}}, \ and\ \bibinfo {author}
  {\bibfnamefont {W.}~\bibnamefont {Richtering}},\ }\href {\doibase
  10.1021/acs.macromol.9b00729} {\bibfield  {journal} {\bibinfo  {journal}
  {Macromol.}\ }\textbf {\bibinfo {volume} {52}},\ \bibinfo {pages} {3995}
  (\bibinfo {year} {2019})}\BibitemShut {NoStop}%
\bibitem [{\citenamefont {Asakura}\ and\ \citenamefont
  {Oosawa}(1954)}]{Asakura1954}%
  \BibitemOpen
  \bibfield  {author} {\bibinfo {author} {\bibfnamefont {S.}~\bibnamefont
  {Asakura}}\ and\ \bibinfo {author} {\bibfnamefont {F.}~\bibnamefont
  {Oosawa}},\ }\href {\doibase 10.1063/1.1740347} {\bibfield  {journal}
  {\bibinfo  {journal} {J. Chem. Phys.}\ }\textbf {\bibinfo {volume} {22}},\
  \bibinfo {pages} {1255} (\bibinfo {year} {1954})}\BibitemShut {NoStop}%
\bibitem [{\citenamefont {Asakura}\ and\ \citenamefont
  {Oosawa}(1958)}]{Asakura1958}%
  \BibitemOpen
  \bibfield  {author} {\bibinfo {author} {\bibfnamefont {S.}~\bibnamefont
  {Asakura}}\ and\ \bibinfo {author} {\bibfnamefont {F.}~\bibnamefont
  {Oosawa}},\ }\href {\doibase 10.1002/pol.1958.1203312618} {\bibfield
  {journal} {\bibinfo  {journal} {J. Polym. Sci.}\ }\textbf {\bibinfo {volume}
  {33}},\ \bibinfo {pages} {183} (\bibinfo {year} {1958})}\BibitemShut
  {NoStop}%
\bibitem [{\citenamefont {Lim}\ and\ \citenamefont
  {Denton}(2016{\natexlab{a}})}]{lim-denton-JCP2016}%
  \BibitemOpen
  \bibfield  {author} {\bibinfo {author} {\bibfnamefont {W.~K.}\ \bibnamefont
  {Lim}}\ and\ \bibinfo {author} {\bibfnamefont {A.~R.}\ \bibnamefont
  {Denton}},\ }\href@noop {} {\bibfield  {journal} {\bibinfo  {journal} {J.
  Chem. Phys.}\ }\textbf {\bibinfo {volume} {144}},\ \bibinfo {pages} {024904}
  (\bibinfo {year} {2016}{\natexlab{a}})}\BibitemShut {NoStop}%
\bibitem [{\citenamefont {Sch{\"a}fer}(1999)}]{schaefer1999}%
  \BibitemOpen
  \bibfield  {author} {\bibinfo {author} {\bibfnamefont {L.}~\bibnamefont
  {Sch{\"a}fer}},\ }\href@noop {} {\emph {\bibinfo {title} {Excluded Volume
  Effects in Polymer Solutions as Explained by the Renormalization Group}}}\
  (\bibinfo  {publisher} {Springer},\ \bibinfo {address} {Berlin},\ \bibinfo
  {year} {1999})\BibitemShut {NoStop}%
\bibitem [{\citenamefont {Saunders}\ and\ \citenamefont
  {Vincent}(1996)}]{Saunders1996}%
  \BibitemOpen
  \bibfield  {author} {\bibinfo {author} {\bibfnamefont {B.~R.}\ \bibnamefont
  {Saunders}}\ and\ \bibinfo {author} {\bibfnamefont {B.}~\bibnamefont
  {Vincent}},\ }\href {\doibase 10.1039/FT9969203385} {\bibfield  {journal}
  {\bibinfo  {journal} {J. Chem. Soc., Faraday Trans.}\ }\textbf {\bibinfo
  {volume} {92}},\ \bibinfo {pages} {3385} (\bibinfo {year}
  {1996})}\BibitemShut {NoStop}%
\bibitem [{\citenamefont {Saunders}\ and\ \citenamefont
  {Vincent}(1997)}]{Saunders1997a}%
  \BibitemOpen
  \bibfield  {author} {\bibinfo {author} {\bibfnamefont {B.~R.}\ \bibnamefont
  {Saunders}}\ and\ \bibinfo {author} {\bibfnamefont {B.}~\bibnamefont
  {Vincent}},\ }\href {\doibase 10.1007/s003960050047} {\bibfield  {journal}
  {\bibinfo  {journal} {Colloid Polym. Sci.}\ }\textbf {\bibinfo {volume}
  {275}},\ \bibinfo {pages} {9} (\bibinfo {year} {1997})}\BibitemShut {NoStop}%
\bibitem [{\citenamefont {Bradley}\ \emph {et~al.}(2005)\citenamefont
  {Bradley}, \citenamefont {Ramos},\ and\ \citenamefont
  {Vincent}}]{bradley-vincent2005}%
  \BibitemOpen
  \bibfield  {author} {\bibinfo {author} {\bibfnamefont {M.}~\bibnamefont
  {Bradley}}, \bibinfo {author} {\bibfnamefont {J.}~\bibnamefont {Ramos}}, \
  and\ \bibinfo {author} {\bibfnamefont {B.}~\bibnamefont {Vincent}},\
  }\href@noop {} {\bibfield  {journal} {\bibinfo  {journal} {Langmuir}\
  }\textbf {\bibinfo {volume} {21}},\ \bibinfo {pages} {1209} (\bibinfo {year}
  {2005})}\BibitemShut {NoStop}%
\bibitem [{\citenamefont {Routh}\ \emph {et~al.}(2006)\citenamefont {Routh},
  \citenamefont {Fern{\'a}ndez-Nieves}, \citenamefont {Bradley},\ and\
  \citenamefont {Vincent}}]{routh2006effect}%
  \BibitemOpen
  \bibfield  {author} {\bibinfo {author} {\bibfnamefont {A.~F.}\ \bibnamefont
  {Routh}}, \bibinfo {author} {\bibfnamefont {A.}~\bibnamefont
  {Fern{\'a}ndez-Nieves}}, \bibinfo {author} {\bibfnamefont {M.}~\bibnamefont
  {Bradley}}, \ and\ \bibinfo {author} {\bibfnamefont {B.}~\bibnamefont
  {Vincent}},\ }\href {\doibase 10.1021/jp0558831} {\bibfield  {journal}
  {\bibinfo  {journal} {J. Phys. Chem. B}\ }\textbf {\bibinfo {volume} {110}},\
  \bibinfo {pages} {12721} (\bibinfo {year} {2006})}\BibitemShut {NoStop}%
\bibitem [{\citenamefont {Fern{\'a}ndez-Nieves}\ \emph
  {et~al.}(2003)\citenamefont {Fern{\'a}ndez-Nieves}, \citenamefont
  {Fern{\'a}ndez-Barbero}, \citenamefont {Vincent},\ and\ \citenamefont {de~las
  Nieves}}]{FernandezNieves2003}%
  \BibitemOpen
  \bibfield  {author} {\bibinfo {author} {\bibfnamefont {A.}~\bibnamefont
  {Fern{\'a}ndez-Nieves}}, \bibinfo {author} {\bibfnamefont {A.}~\bibnamefont
  {Fern{\'a}ndez-Barbero}}, \bibinfo {author} {\bibfnamefont {B.}~\bibnamefont
  {Vincent}}, \ and\ \bibinfo {author} {\bibfnamefont {F.~J.}\ \bibnamefont
  {de~las Nieves}},\ }\href {\doibase 10.1063/1.1618734} {\bibfield  {journal}
  {\bibinfo  {journal} {J. Chem. Phys.}\ }\textbf {\bibinfo {volume} {119}},\
  \bibinfo {pages} {10383} (\bibinfo {year} {2003})}\BibitemShut {NoStop}%
\bibitem [{\citenamefont {Bastide}\ \emph {et~al.}(1981)\citenamefont
  {Bastide}, \citenamefont {Candau},\ and\ \citenamefont
  {Leibler}}]{Bastide1981}%
  \BibitemOpen
  \bibfield  {author} {\bibinfo {author} {\bibfnamefont {J.}~\bibnamefont
  {Bastide}}, \bibinfo {author} {\bibfnamefont {S.}~\bibnamefont {Candau}}, \
  and\ \bibinfo {author} {\bibfnamefont {L.}~\bibnamefont {Leibler}},\ }\href
  {\doibase 10.1021/ma50004a050} {\bibfield  {journal} {\bibinfo  {journal}
  {Macromol.}\ }\textbf {\bibinfo {volume} {14}},\ \bibinfo {pages} {719}
  (\bibinfo {year} {1981})}\BibitemShut {NoStop}%
\bibitem [{\citenamefont {Benhamou}\ \emph {et~al.}(2016)\citenamefont
  {Benhamou}, \citenamefont {Moussa},\ and\ \citenamefont
  {Kaidi}}]{Benhamou2016}%
  \BibitemOpen
  \bibfield  {author} {\bibinfo {author} {\bibfnamefont {M.}~\bibnamefont
  {Benhamou}}, \bibinfo {author} {\bibfnamefont {F.~E.}\ \bibnamefont
  {Moussa}}, \ and\ \bibinfo {author} {\bibfnamefont {H.}~\bibnamefont
  {Kaidi}},\ }\href {\doibase 10.3390/polym8040121} {\bibfield  {journal}
  {\bibinfo  {journal} {Polymers}\ }\textbf {\bibinfo {volume} {8}},\ \bibinfo
  {pages} {121} (\bibinfo {year} {2016})}\BibitemShut {NoStop}%
\bibitem [{\citenamefont {Molina}\ \emph {et~al.}(2005)\citenamefont {Molina},
  \citenamefont {Uribe-Arocha},\ and\ \citenamefont {Osswald}}]{Molina2005}%
  \BibitemOpen
  \bibfield  {author} {\bibinfo {author} {\bibfnamefont {R.}~\bibnamefont
  {Molina}}, \bibinfo {author} {\bibfnamefont {J.~A.}\ \bibnamefont
  {Uribe-Arocha}}, \ and\ \bibinfo {author} {\bibfnamefont {T.~A.}\
  \bibnamefont {Osswald}},\ }\href {\doibase 10.1021/ma047763e} {\bibfield
  {journal} {\bibinfo  {journal} {Macromol.}\ }\textbf {\bibinfo {volume}
  {38}},\ \bibinfo {pages} {4219} (\bibinfo {year} {2005})}\BibitemShut
  {NoStop}%
\bibitem [{\citenamefont {Takarai}\ \emph {et~al.}(2026)\citenamefont
  {Takarai}, \citenamefont {Yasuda}, \citenamefont {Sakumichi},\ and\
  \citenamefont {Sakai}}]{Takarai2026}%
  \BibitemOpen
  \bibfield  {author} {\bibinfo {author} {\bibfnamefont {H.}~\bibnamefont
  {Takarai}}, \bibinfo {author} {\bibfnamefont {T.}~\bibnamefont {Yasuda}},
  \bibinfo {author} {\bibfnamefont {N.}~\bibnamefont {Sakumichi}}, \ and\
  \bibinfo {author} {\bibfnamefont {T.}~\bibnamefont {Sakai}},\ }\href@noop {}
  {\bibfield  {journal} {\bibinfo  {journal} {Phys. Rev. E}\ }\textbf {\bibinfo
  {volume} {114}},\ \bibinfo {pages} {015420} (\bibinfo {year}
  {2026})}\BibitemShut {NoStop}%
\bibitem [{\citenamefont {Aryal}\ and\ \citenamefont
  {Denton}(2026)}]{aryal2026response}%
  \BibitemOpen
  \bibfield  {author} {\bibinfo {author} {\bibfnamefont {M.}~\bibnamefont
  {Aryal}}\ and\ \bibinfo {author} {\bibfnamefont {A.~R.}\ \bibnamefont
  {Denton}},\ }\href@noop {} {\bibfield  {journal} {\bibinfo  {journal} {J.
  Chem. Phys.}\ }\textbf {\bibinfo {volume} {164}},\ \bibinfo {pages} {064904}
  (\bibinfo {year} {2026})}\BibitemShut {NoStop}%
\bibitem [{\citenamefont {de~Gennes}(1979)}]{deGennes1979}%
  \BibitemOpen
  \bibfield  {author} {\bibinfo {author} {\bibfnamefont {P.-G.}\ \bibnamefont
  {de~Gennes}},\ }\href@noop {} {\emph {\bibinfo {title} {Scaling Concepts in
  Polymer Physics}}}\ (\bibinfo  {publisher} {Cornell},\ \bibinfo {address}
  {Ithaca},\ \bibinfo {year} {1979})\BibitemShut {NoStop}%
\bibitem [{\citenamefont {Rubinstein}\ and\ \citenamefont
  {Colby}(2003)}]{rubinstein2003polymer}%
  \BibitemOpen
  \bibfield  {author} {\bibinfo {author} {\bibfnamefont {M.}~\bibnamefont
  {Rubinstein}}\ and\ \bibinfo {author} {\bibfnamefont {R.~H.}\ \bibnamefont
  {Colby}},\ }\href@noop {} {\emph {\bibinfo {title} {Polymer physics}}}\
  (\bibinfo  {publisher} {Oxford university press},\ \bibinfo {year}
  {2003})\BibitemShut {NoStop}%
\bibitem [{\citenamefont {Doi}(2013)}]{Doi2013}%
  \BibitemOpen
  \bibfield  {author} {\bibinfo {author} {\bibfnamefont {M.}~\bibnamefont
  {Doi}},\ }\href@noop {} {\emph {\bibinfo {title} {Soft Matter Physics}}}\
  (\bibinfo  {publisher} {Oxford University Press},\ \bibinfo {address}
  {Oxford},\ \bibinfo {year} {2013})\BibitemShut {NoStop}%
\bibitem [{\citenamefont {Dhar}(1978)}]{Dhar1978}%
  \BibitemOpen
  \bibfield  {author} {\bibinfo {author} {\bibfnamefont {D.}~\bibnamefont
  {Dhar}},\ }\href {\doibase 10.1063/1.523515} {\bibfield  {journal} {\bibinfo
  {journal} {J. Math. Phys.}\ }\textbf {\bibinfo {volume} {19}},\ \bibinfo
  {pages} {5} (\bibinfo {year} {1978})}\BibitemShut {NoStop}%
\bibitem [{\citenamefont {Kuhn}(1934)}]{kuhn1934shape}%
  \BibitemOpen
  \bibfield  {author} {\bibinfo {author} {\bibfnamefont {W.}~\bibnamefont
  {Kuhn}},\ }\href@noop {} {\bibfield  {journal} {\bibinfo  {journal}
  {Kolloid-Zeitschrift}\ }\textbf {\bibinfo {volume} {68}} (\bibinfo {year}
  {1934})}\BibitemShut {NoStop}%
\bibitem [{\citenamefont {Theodorou}\ and\ \citenamefont
  {Suter}(1985)}]{Theodorou1985}%
  \BibitemOpen
  \bibfield  {author} {\bibinfo {author} {\bibfnamefont {D.~N.}\ \bibnamefont
  {Theodorou}}\ and\ \bibinfo {author} {\bibfnamefont {U.~W.}\ \bibnamefont
  {Suter}},\ }\href {\doibase 10.1021/ma00148a028} {\bibfield  {journal}
  {\bibinfo  {journal} {Macromol.}\ }\textbf {\bibinfo {volume} {18}},\
  \bibinfo {pages} {1206} (\bibinfo {year} {1985})}\BibitemShut {NoStop}%
\bibitem [{\citenamefont {Rudnick}\ and\ \citenamefont
  {Gaspari}(1987)}]{rudnick1987shapes}%
  \BibitemOpen
  \bibfield  {author} {\bibinfo {author} {\bibfnamefont {J.}~\bibnamefont
  {Rudnick}}\ and\ \bibinfo {author} {\bibfnamefont {G.}~\bibnamefont
  {Gaspari}},\ }\href@noop {} {\bibfield  {journal} {\bibinfo  {journal}
  {Science}\ }\textbf {\bibinfo {volume} {237}},\ \bibinfo {pages} {384}
  (\bibinfo {year} {1987})}\BibitemShut {NoStop}%
\bibitem [{\citenamefont {Arkın}\ and\ \citenamefont
  {Janke}(2013)}]{Arkin2013}%
  \BibitemOpen
  \bibfield  {author} {\bibinfo {author} {\bibfnamefont {H.}~\bibnamefont
  {Arkın}}\ and\ \bibinfo {author} {\bibfnamefont {W.}~\bibnamefont {Janke}},\
  }\href {\doibase 10.1063/1.4788616} {\bibfield  {journal} {\bibinfo
  {journal} {J. Chem. Phys.}\ }\textbf {\bibinfo {volume} {138}},\ \bibinfo
  {pages} {054904} (\bibinfo {year} {2013})}\BibitemShut {NoStop}%
\bibitem [{\citenamefont {Sciutto}(1996)}]{sciutto1996}%
  \BibitemOpen
  \bibfield  {author} {\bibinfo {author} {\bibfnamefont {S.~J.}\ \bibnamefont
  {Sciutto}},\ }\href@noop {} {\bibfield  {journal} {\bibinfo  {journal} {J.
  Phys. A: Math. Gen.}\ }\textbf {\bibinfo {volume} {29}},\ \bibinfo {pages}
  {5455} (\bibinfo {year} {1996})}\BibitemShut {NoStop}%
\bibitem [{\citenamefont {Lim}\ and\ \citenamefont
  {Denton}(2014)}]{lim-denton-JCP2014}%
  \BibitemOpen
  \bibfield  {author} {\bibinfo {author} {\bibfnamefont {W.~K.}\ \bibnamefont
  {Lim}}\ and\ \bibinfo {author} {\bibfnamefont {A.~R.}\ \bibnamefont
  {Denton}},\ }\href@noop {} {\bibfield  {journal} {\bibinfo  {journal} {J.
  Chem. Phys.}\ }\textbf {\bibinfo {volume} {141}},\ \bibinfo {pages} {114909}
  (\bibinfo {year} {2014})}\BibitemShut {NoStop}%
\bibitem [{\citenamefont {Lim}\ and\ \citenamefont
  {Denton}(2016{\natexlab{b}})}]{lim-denton-SM2016}%
  \BibitemOpen
  \bibfield  {author} {\bibinfo {author} {\bibfnamefont {W.~K.}\ \bibnamefont
  {Lim}}\ and\ \bibinfo {author} {\bibfnamefont {A.~R.}\ \bibnamefont
  {Denton}},\ }\href@noop {} {\bibfield  {journal} {\bibinfo  {journal} {Soft
  Matter}\ }\textbf {\bibinfo {volume} {12}},\ \bibinfo {pages} {2247}
  (\bibinfo {year} {2016}{\natexlab{b}})}\BibitemShut {NoStop}%
\bibitem [{\citenamefont {Davis}\ and\ \citenamefont
  {Denton}(2018)}]{davis-denton-JCP2018}%
  \BibitemOpen
  \bibfield  {author} {\bibinfo {author} {\bibfnamefont {W.~J.}\ \bibnamefont
  {Davis}}\ and\ \bibinfo {author} {\bibfnamefont {A.~R.}\ \bibnamefont
  {Denton}},\ }\href@noop {} {\bibfield  {journal} {\bibinfo  {journal} {J.
  Chem. Phys.}\ }\textbf {\bibinfo {volume} {149}},\ \bibinfo {pages} {124901}
  (\bibinfo {year} {2018})}\BibitemShut {NoStop}%
\bibitem [{\citenamefont {Lekkerkerker}\ \emph {et~al.}(1992)\citenamefont
  {Lekkerkerker}, \citenamefont {Poon}, \citenamefont {Pusey}, \citenamefont
  {Stroobants},\ and\ \citenamefont {Warren}}]{lekkerkerker1992}%
  \BibitemOpen
  \bibfield  {author} {\bibinfo {author} {\bibfnamefont {H.~N.~W.}\
  \bibnamefont {Lekkerkerker}}, \bibinfo {author} {\bibfnamefont {W.~C.~K.}\
  \bibnamefont {Poon}}, \bibinfo {author} {\bibfnamefont {P.~N.}\ \bibnamefont
  {Pusey}}, \bibinfo {author} {\bibfnamefont {A.}~\bibnamefont {Stroobants}}, \
  and\ \bibinfo {author} {\bibfnamefont {P.~B.}\ \bibnamefont {Warren}},\
  }\href@noop {} {\bibfield  {journal} {\bibinfo  {journal} {Europhys. Lett.}\
  }\textbf {\bibinfo {volume} {20}},\ \bibinfo {pages} {559} (\bibinfo {year}
  {1992})}\BibitemShut {NoStop}%
\bibitem [{\citenamefont {Denton}\ and\ \citenamefont
  {Davis}(2021)}]{denton-davis-JCP2021}%
  \BibitemOpen
  \bibfield  {author} {\bibinfo {author} {\bibfnamefont {A.~R.}\ \bibnamefont
  {Denton}}\ and\ \bibinfo {author} {\bibfnamefont {W.~J.}\ \bibnamefont
  {Davis}},\ }\href@noop {} {\bibfield  {journal} {\bibinfo  {journal} {J.
  Chem. Phys.}\ }\textbf {\bibinfo {volume} {155}},\ \bibinfo {pages} {084904}
  (\bibinfo {year} {2021})}\BibitemShut {NoStop}%
\bibitem [{\citenamefont {Briber}\ \emph {et~al.}(1995)\citenamefont {Briber},
  \citenamefont {Liu},\ and\ \citenamefont {Bauer}}]{briber1995}%
  \BibitemOpen
  \bibfield  {author} {\bibinfo {author} {\bibfnamefont {R.~M.}\ \bibnamefont
  {Briber}}, \bibinfo {author} {\bibfnamefont {X.}~\bibnamefont {Liu}}, \ and\
  \bibinfo {author} {\bibfnamefont {B.~J.}\ \bibnamefont {Bauer}},\ }\href@noop
  {} {\bibfield  {journal} {\bibinfo  {journal} {Science}\ }\textbf {\bibinfo
  {volume} {268}},\ \bibinfo {pages} {395} (\bibinfo {year}
  {1995})}\BibitemShut {NoStop}%
\bibitem [{\citenamefont {Liu}\ \emph {et~al.}(1997)\citenamefont {Liu},
  \citenamefont {Bauer},\ and\ \citenamefont {Briber}}]{briber1997}%
  \BibitemOpen
  \bibfield  {author} {\bibinfo {author} {\bibfnamefont {X.}~\bibnamefont
  {Liu}}, \bibinfo {author} {\bibfnamefont {B.~J.}\ \bibnamefont {Bauer}}, \
  and\ \bibinfo {author} {\bibfnamefont {R.~M.}\ \bibnamefont {Briber}},\
  }\href@noop {} {\bibfield  {journal} {\bibinfo  {journal} {Macromol.}\
  }\textbf {\bibinfo {volume} {30}},\ \bibinfo {pages} {4704} (\bibinfo {year}
  {1997})}\BibitemShut {NoStop}%
\bibitem [{\citenamefont {Honeycutt}\ and\ \citenamefont
  {Thirumalai}(1989)}]{honeycutt1989}%
  \BibitemOpen
  \bibfield  {author} {\bibinfo {author} {\bibfnamefont {J.~D.}\ \bibnamefont
  {Honeycutt}}\ and\ \bibinfo {author} {\bibfnamefont {D.}~\bibnamefont
  {Thirumalai}},\ }\href@noop {} {\bibfield  {journal} {\bibinfo  {journal} {J.
  Chem. Phys.}\ }\textbf {\bibinfo {volume} {90}},\ \bibinfo {pages} {4542}
  (\bibinfo {year} {1989})}\BibitemShut {NoStop}%
\bibitem [{\citenamefont {Levin}(2002)}]{levin2002}%
  \BibitemOpen
  \bibfield  {author} {\bibinfo {author} {\bibfnamefont {Y.}~\bibnamefont
  {Levin}},\ }\href@noop {} {\bibfield  {journal} {\bibinfo  {journal} {Phys.
  Rev. E}\ }\textbf {\bibinfo {volume} {65}},\ \bibinfo {pages} {036143}
  (\bibinfo {year} {2002})}\BibitemShut {NoStop}%
\bibitem [{\citenamefont {Colla}\ \emph {et~al.}(2014)\citenamefont {Colla},
  \citenamefont {Likos},\ and\ \citenamefont {Levin}}]{colla-likos2014}%
  \BibitemOpen
  \bibfield  {author} {\bibinfo {author} {\bibfnamefont {T.}~\bibnamefont
  {Colla}}, \bibinfo {author} {\bibfnamefont {C.~N.}\ \bibnamefont {Likos}}, \
  and\ \bibinfo {author} {\bibfnamefont {Y.}~\bibnamefont {Levin}},\
  }\href@noop {} {\bibfield  {journal} {\bibinfo  {journal} {J. Chem. Phys.}\
  }\textbf {\bibinfo {volume} {141}},\ \bibinfo {pages} {234902} (\bibinfo
  {year} {2014})}\BibitemShut {NoStop}%
\bibitem [{\citenamefont {Moncho-Jord{\'a}}(2013)}]{moncho-jorda2013}%
  \BibitemOpen
  \bibfield  {author} {\bibinfo {author} {\bibfnamefont {A.}~\bibnamefont
  {Moncho-Jord{\'a}}},\ }\href@noop {} {\bibfield  {journal} {\bibinfo
  {journal} {J. Chem. Phys.}\ }\textbf {\bibinfo {volume} {139}},\ \bibinfo
  {pages} {064906} (\bibinfo {year} {2013})}\BibitemShut {NoStop}%
\bibitem [{\citenamefont {Adroher-Ben{\'i}tez}\ \emph
  {et~al.}(2015)\citenamefont {Adroher-Ben{\'i}tez}, \citenamefont {Ahualli},
  \citenamefont {Mart{\'i}n-Molina}, \citenamefont {Quesada-P{\'e}rez},\ and\
  \citenamefont {Moncho-Jord{\'a}}}]{quesada-perez-moncho-jorda-anta2015}%
  \BibitemOpen
  \bibfield  {author} {\bibinfo {author} {\bibfnamefont {I.}~\bibnamefont
  {Adroher-Ben{\'i}tez}}, \bibinfo {author} {\bibfnamefont {S.}~\bibnamefont
  {Ahualli}}, \bibinfo {author} {\bibfnamefont {A.}~\bibnamefont
  {Mart{\'i}n-Molina}}, \bibinfo {author} {\bibfnamefont {M.}~\bibnamefont
  {Quesada-P{\'e}rez}}, \ and\ \bibinfo {author} {\bibfnamefont
  {A.}~\bibnamefont {Moncho-Jord{\'a}}},\ }\href@noop {} {\bibfield  {journal}
  {\bibinfo  {journal} {Macromol.}\ }\textbf {\bibinfo {volume} {48}},\
  \bibinfo {pages} {4645} (\bibinfo {year} {2015})}\BibitemShut {NoStop}%
\bibitem [{\citenamefont {Moncho-Jord{\'a}}\ and\ \citenamefont
  {Dzubiella}(2016)}]{moncho-jorda-dzubiella2016}%
  \BibitemOpen
  \bibfield  {author} {\bibinfo {author} {\bibfnamefont {A.}~\bibnamefont
  {Moncho-Jord{\'a}}}\ and\ \bibinfo {author} {\bibfnamefont {J.}~\bibnamefont
  {Dzubiella}},\ }\href@noop {} {\bibfield  {journal} {\bibinfo  {journal}
  {Phys. Chem. Chem. Phys.}\ }\textbf {\bibinfo {volume} {18}},\ \bibinfo
  {pages} {5372} (\bibinfo {year} {2016})}\BibitemShut {NoStop}%
\end{thebibliography}%

\end{document}